\documentclass[aps,prx,superscriptaddress,twocolumn,longbibliography]{revtex4-2}
\usepackage{amssymb}
\usepackage{amsmath}
\usepackage{graphicx}
\usepackage{dcolumn}
\usepackage{epsfig}
\usepackage{float}
\usepackage{changes}
\usepackage{comment}
\usepackage[bbgreekl]{mathbbol}

\usepackage{mathrsfs}

\usepackage[cal=boondox,scr=boondoxo]{mathalfa}

\usepackage[colorlinks=true,linkcolor=blue,urlcolor=black,citecolor=blue,bookmarksopen=true]{hyperref}

\usepackage{trsym} 

\usepackage[colorlinks=true,linkcolor=blue,urlcolor=black,citecolor=blue,bookmarksopen=true]{hyperref}

\usepackage{graphicx}
\usepackage{dcolumn}
\usepackage{epsfig}

\usepackage[bbgreekl]{mathbbol}

\usepackage{soul}

\def\bra#1{\mathinner{\langle{#1}|}}
\def\ket#1{\mathinner{|{#1}\rangle}}

\def\braket#1{\mathinner{\langle{#1}\rangle}}
\def\sandwich#1#2#3{\mathinner{\langle{#1}|{#2}|{#3}\rangle}}

\def\bbraket#1{\mathinner{\langle\hspace{-0.75mm}\langle{#1}\rangle\hspace{-0.75mm}\rangle}}

\def\Tr{\mathrm{Tr}}

\def\I{\mathrm{I}}
\usepackage[T3,T1]{fontenc}
\DeclareSymbolFont{tipa}{T3}{cmr}{m}{n}
\DeclareMathAccent{\invbreve}{\mathalpha}{tipa}{16}

\usepackage{accents}
\newlength{\hhatheight}

\usepackage{xcolor}
\newcommand{\extra}[1]{}

\makeatother

\begin{document}

\title{Hilbert Space Fragmentation and Subspace Scar Time-Crystallinity in Driven Homogeneous Central-Spin Models}

\author{Abhishek Kumar}
\affiliation{Department of Physics, Virginia Tech, Blacksburg, VA 24061, USA}
\affiliation{Virginia Tech Center for Quantum Information Science and Engineering, Blacksburg, VA 24061, USA}
\author{Rafail Frantzeskakis}
\affiliation{Department of Physics, Virginia Tech, Blacksburg, VA 24061, USA}
\affiliation{Virginia Tech Center for Quantum Information Science and Engineering, Blacksburg, VA 24061, USA}
\affiliation{Department of Physics, University of Crete, Heraklion, 71003, Greece}

\author{Edwin Barnes}
\affiliation{Department of Physics, Virginia Tech, Blacksburg, VA 24061, USA}
\affiliation{Virginia Tech Center for Quantum Information Science and Engineering, Blacksburg, VA 24061, USA}

\begin{abstract}
We study the stroboscopic non-equilibrium quantum dynamics of periodically kicked Hamiltonians involving homogeneous central-spin interactions. The system exhibits a strong fragmentation of Hilbert space into four-dimensional Floquet-Krylov subspaces, which oscillate between two disjointed two-dimensional subspaces and thus break the discrete time-translation symmetry of the system. Our analytical and numerical analyses reveal that fully polarized states of the satellite spins exhibit fragmentations that are stable against perturbations and have high overlap with Floquet eigenstates of atypically low bipartite entanglement entropy (scar states). Motivated by the breaking of discrete time translation symmetry by Floquet-Krylov subspaces, we introduce a novel type of time crystal that we call a ``subspace time crystal''.
We present evidence of robust time-crystalline behavior in the form of a period doubling of the total magnetization of fully polarized satellite spin states that persists over long time scales. We compute non-equilibrium phase diagrams with respect to a magnetic field, coupling terms, and pulse error for various interaction types, including Heisenberg, Ising, XXZ, and XX. We also discuss possible experimental realizations of scar time crystals in color center, quantum dot, and rare-earth ion platforms.  
\end{abstract}

\date{\today}

{
\let\clearpage\relax
\maketitle
}

\section{Introduction }
Quantum discrete time crystals (DTCs) are non-equilibrium phases of matter that can only exist in periodically driven many-body systems with short-range interactions \cite{else2016floquet,khemani2019brief,else2020discrete,zaletel2023colloquium}.  In a DTC,  a non-equilibrium initial state breaks the discrete time-translation symmetry of the Hamiltonian, which can be diagnosed by an expectation value of a local observable evolving subharmonically relative to the driving period in the long-time limit~\cite{watanabe2015absence,khemani2019brief}. Non-equilibrium quantum states, particularly in the DTC phase, must break ergodicity \footnote{Broadly, ergodicity refers to the phenomenon where the time average of the expectation value of an observable is equal to the ensemble (density matrix) average.} so that the expectation value of an operator does not converge to one value at long times. A generic (interacting and non-integrable) isolated many-body quantum system typically expresses ergodic behavior through the convergence of its reduced density matrices to thermal states, as postulated in the eigenstate thermalization hypothesis (ETH)\cite{kim2014testing,alba2015eigenstate,deutsch2018eigenstate}.

In most DTCs, the breaking of ergodicity has been attributed to many-body localization (MBL) induced by random disorder \cite{pal2010many,vosk2015theory,choi2016exploring,abanin2019colloquium}. In the strong breaking of ergodicity, all eigenstates are non-thermal (they do not obey the ETH and have atypically low entanglement entropy) and all product states behave like non-equilibrium states for some local observable. However, ergodicity can also be weakly broken through the quantum many-body scar phenomenon  \cite{choi2019emergent,yao2022quantum,you2022quantum,bocini2022growing,desaules2023weak,desaules2023prominent,mizuta2020exact,sugiura2021many,bhattacharjee2022probing,windt2022squeezing,moudgalya2020eta,desaules2021proposal,hummel2022genuine}, which can manifest in several ways \cite{Chandran_ARCP_2023_Scar}: (i) There is a tiny fraction of non-thermal eigenstates that have atypically low bipartite entanglement entropy; (ii) There is a coherent revival of some initial many-body states; (iii) The system possesses some species of infinitely long-lived quasiparticles. Additionally, a conceptually distinct but related phenomenon of weak ergodicity breaking is known as Hilbert space fragmentation (HSF) \cite{Sala2020Hilbert_Frag, moudgalya2020eta,Moudgalya_2022_Fragmentation} or Hilbert space shattering \cite{Khemani_2020_Hilbert_Shatter},  where Hilbert space can be decomposed into an exponentially large number of dynamically disconnected sectors. HSF mostly occurs in Hamiltonian systems with global symmetries or in random unitary circuits \cite{Moudgalya_2022_Fragmentation}. It is also regarded as part of the quantum scar phenomenon, as its ergodicity breaking \cite{Chandran_ARCP_2023_Scar} depends on the initial state.

Scar time-crystallinity has been observed in one- and two-dimensional lattices on a quantum simulator based on Rydberg atom arrays \cite{Bluvstein_2021_ScarTC} and theoretically studied using the PXP model \cite{Maskara_PRL_ScarTC2021}.  Here, a periodic drive stabilizes coherent revivals of quantum many-body scars against perturbations, thus leading to a time-crystalline behavior for N\'eel-like states. Another study by Huang et al. \cite{Huang_2022_ScarTC} explores scar time crystallinity in a bosonic model, wherein quantum scars are maintained via spatiotemporal symmetries. Unlike a conventional (MBL-type) time crystal, a scar time crystal exhibits a robust subharmonic response only for a few initial states having large overlaps to the scar states.

Most of the studies on MBL time crystals focus on spin models with Ising-type interactions, while work on scar time crystals has largely centered on the PXP model. Exploring new classes of interacting many-body systems that exhibit time-crystalline behavior could not only provide additional insight into the breaking of ergodicity, but could also help identify new platforms for  experimental realizations that are closer to the thermodynamic limit, and which could thus potentially address outstanding conceptual issues related to the existence of such phases~\cite{Morningstar_PRB2022}. In this regard, central-spin models offer significant opportunities due to their realization on various platforms like color centers~\cite{dreau2014probing,rose2018observation,soltamov2021electron} and quantum dots ~\cite{sallen2014nuclear,chekhovich2017measurement,denning2019collective,jackson2021quantum}, which can contain up to a few million spins.  
Time-crystalline behavior in Ising-type central-spin systems has been analyzed and experimentally studied in the NMR setting~\cite{Pal_PRL_2018}. Time-crystallinity in systems with Heisenberg interactions, such as in quantum dot arrays, has been studied theoretically as well~\cite{Barnes_2019_HeisenTC, Li_2020_HeisenTC}. More recently, two of the present authors showed that time-crystalline behavior can emerge in Heisenberg and XXZ central-spin models with random coupling disorder when there is a large Zeeman energy mismatch between central and satellite spins or when additional pulses are applied to the central spin every Floquet period.

When considering XXZ central-spin models with homogeneous couplings, certain global symmetries that are absent in the case of random coupling disorder emerge. The presence of global symmetries in a quantum system can sometimes lead to HSF~\cite{Sala2020Hilbert_Frag,Moudgalya_2022_Fragmentation}, which motivates us to investigate HSF in central-spin systems. Furthermore, the application of a periodic drive may result in stable time-crystallinity for specific initial states, suggesting the possibility of obtaining scar time crystals in homogeneous central-spin models.

In this paper, we show that Hilbert space fragmentation and scar time-crystallinity can occur in a driven, homogeneous central-spin model. 
 We consider a system with homogeneous interactions (of Heisenberg, Ising, XXZ, or XX type) subject to periodic $\pi$-pulse driving.  Within a given symmetry sector of the satellite spin Hilbert space, we find that the stroboscopic dynamics are constrained to a four-dimensional Floquet-Krylov subspace, leading to a fragmentation of Hilbert space. The Floquet-Krylov subspace oscillates between two disjointed regions, breaking the discrete time-translation symmetry of the system. 
The fully polarized satellite states exhibit a more robust fragmentation compared to other initial states, and they have a high overlap with Floquet scar states---the Floquet eigenstates that have low bipartite entanglement entropy. Furthermore, we define ``subspace time crystals'' using the concept of Floquet Krylov subspaces. We diagnose the scar time-crystallinity using total satellite spin magnetization. We note that, in this paper, scar time-crystalline behavior is identified based on the long-lived  (but possibly finite) period doubling that requires interactions and is robust to pulse errors for a finite number of satellite spins---we do not consider the thermodynamic limit, where fundamental conceptual questions about the nature of non-equilibrium phases remain unclear~\cite{Morningstar_PRB2022}.
We show that the stability of the fragmentation and time-crystalline behavior against pulse errors depends on the initial state, the nature of the interaction, and the magnetic field on the central spin. We find that the stability is the most prevalent for Ising interactions, followed by XXZ, Heisenberg, and XX interactions, respectively. Numerical simulations reveal a robust subharmonicity (period doubling) in the satellite magnetization for fully polarized satellite states. We obtain non-equilibrium phase diagrams with respect to pulse error, interaction strength, and magnetic field for different types of interactions.

We have organized the paper as follows.
In Sec.~\ref{sec:Model}, we introduce the homogeneous XXZ central-spin model driven by periodic $\pi$ pulses and study the relevant symmetry of the Floquet Hamiltonian.  We also study the Hilbert space fragmentation and its stability. We further calculate the bipartite entanglement entropy of the satellite part of a Floquet eigenstate and its overlap with fully polarized satellite spin states.
In Sec.~\ref{sec:Scar_TC}, we formally define subspace time crystals and focus on scar time-crystallinity in the system. To diagnose the time-crystalline behavior, we calculate the satellite magnetization numerically. Period doubling of the satellite magnetization is shown by plotting it with respect to various parameters like the magnetic field, coupling terms, and pulse error.
In Sec. \ref{sec:Dynamical_probe_and_Experiment}, we also discuss the experimental realization of time crystals in color center, rare-earth ion, and quantum dot platforms. We conclude in Sec. \ref{sec:Discussion}.
Some technical details are given in the appendices.

 \section{Hilbert Space Fragmentation and Quantum scar in a central spin system}\label{sec:Model}
 In this section,  we introduce the driven homogeneous central-spin model under study in this work, investigate the constrained stroboscopic dynamics in a symmetry sector, and discuss the quantum scar-like behavior of fully polarized states.
\begin{figure}[t]
   \centering
   \includegraphics[width=\columnwidth]{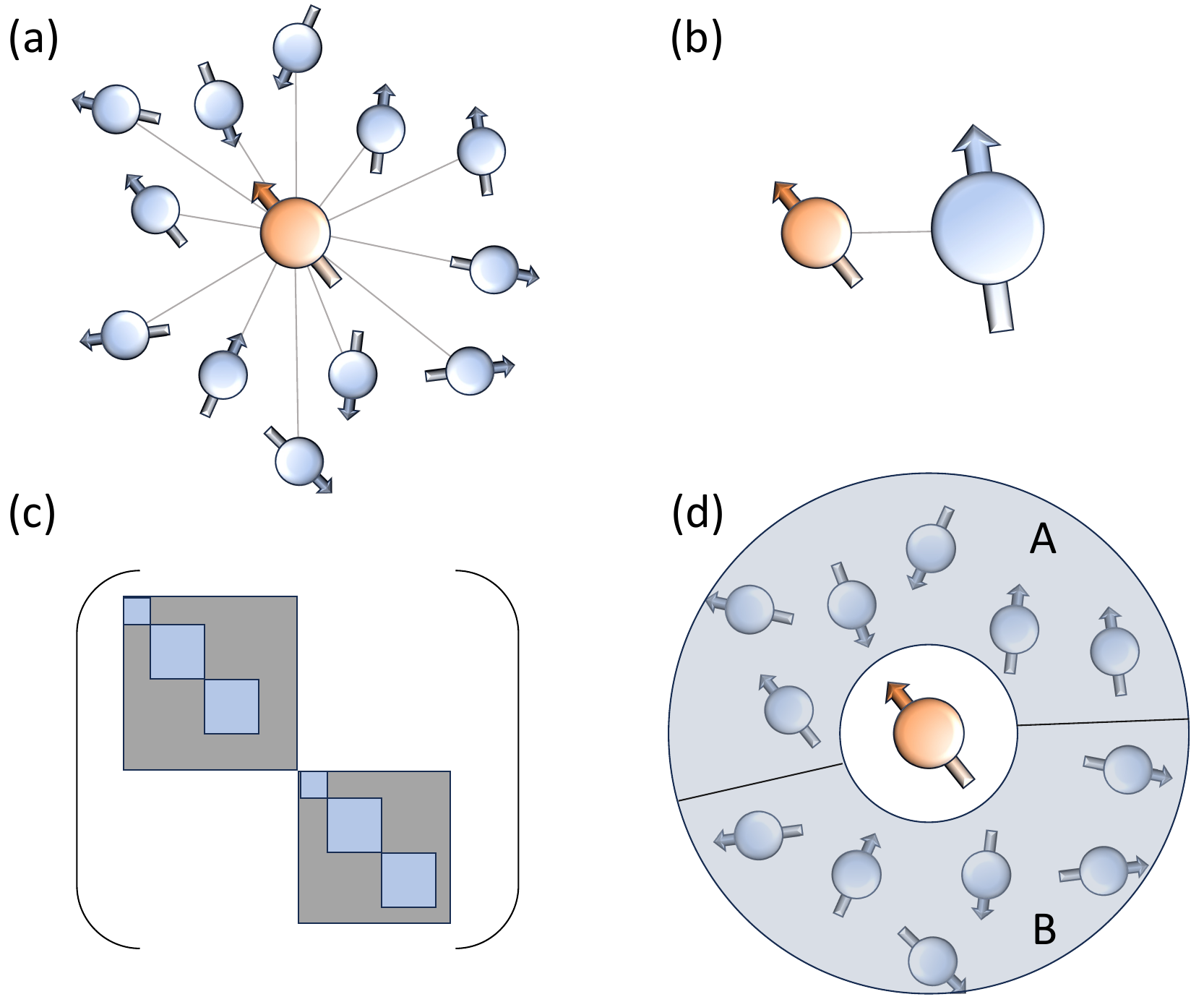}
  \caption{(a) A central-spin model in which the central spin (orange) interacts with all satellite spins (blue) with equal interaction strength. (b) In the case of equal interaction strengths, the satellite spins can be replaced by one large collective satellite spin. (c) The conservation of total satellite spin causes the Floquet operator to block-diagonalize into symmetry sectors (gray boxes). Hilbert-space fragmentation within each symmetry sector leads to a more fine-grained block diagonalization (blue boxes). (d) Bipartition of the satellite spins into two equal parts $A$ and $B$.}
   \label{fig:Fig1_Schematic_Diag}
\end{figure}
\subsection{Model Hamiltonian} 
We consider a central-spin model consisting of one central spin coupled to $N$ satellite spins via XXZ-type interactions with uniform coupling strengths (see Fig.~\ref{fig:Fig1_Schematic_Diag}(a)). In the context of color centers, rare earth ions, or quantum dots, the central spin is an electronic spin that couples to nuclear (satellite) spins via hyperfine interactions. The form of such interactions can vary from Heisenberg to Ising depending on the details of the system in question. A homogeneous central-spin system can effectively represent a dense ensemble of nuclei in a quantum dot. Such systems have been studied in experiments, as discussed in \cite{Zaporski_PRX_Quantum2024_Desnse}. Additionally, homogeneous Heisenberg couplings can describe a GaAs quantum dot realistically at short times and low magnetic fields \cite{Cywiński_PRL_2009,Cywiński_PRB_2009,barnes2011master}. The XXZ model encompasses all of these possibilities.
We can write the Hamiltonian as
\begin{align}\label{eq:Ht}
H_0 &= \sum_{p=1}^{N} \Big[ A_{xy}( \I_{p}^{x}S_{0}^{x} + \I_{p}^{y}S_{0}^{y}) + A_{z} \I_{p}^{z}S_{0}^{z} + B^{S}_{\mu}\I_{p}^{\mu}\Big] \nonumber \\  &\hspace{25mm} + B^{C}_{\mu} S_{0}^{\mu}.
\end{align}
Here, $S^{\mu}_0= \sigma^\mu_0/2$ $(\I_p^{\mu}=\sigma_p^\mu/2)$ denotes a central (satellite) spin operator with $\mu\in \{x,y,z\}$. The parameters $A_{xy}, A_z $ are central-satellite spin coupling constants, while $B^{S}_\mu$ and $B^{C}_\mu$ are Zeeman field components for the satellite and central spins, respectively. We take $B^S_\mu=0$ under the assumption that the satellite spin $g$-factor is orders of magnitude smaller than that of the central spin, which is the case for electronic and nuclear spins in color centers, rare earth ions, and quantum dots. Without loss of generality, we take the Zeeman field on the electron to be in the $z$-direction: $B^{C}_{\mu}=B_z$.

The $z$-component of the total spin of the system is a conserved quantity, $\big[H_0, S_0^z+ \sum_{p=1}^{N} \I_p^z\big]=0$, when the Zeeman fields are in the $z$ direction. The magnitude of the total satellite spin is also conserved, $\big[H_0, \I_t^2\big]=0$, where
 \begin{equation}
 \I_t^2= (\I_t^x)^2+(\I_t^y)^2+(\I_t^z)^2,
 \end{equation}
with $\I_t^{\mu}= \sum_{p=1}^{N} \I_{p}^{\mu}$. We can then replace the individual satellite spins with a single large spin (see Fig.~\ref{fig:Fig1_Schematic_Diag}(b)).
When the spins are periodically driven by instantaneous pulses that implement rotations about the $x$ axis, the total Hamiltonian  becomes $H(t)= H_0 + H_p(t)$, with
\begin{align}
H_0 &= A_{xy}( \I_t^{+}S_{0}^{-}+\I_t^{-}S_{0}^{+}) + A_{z}\I_t^{z}S_{0}^z + B_z S_{0}^z, \label{eq:H0}
\\
H_{p}(t) &= \sum_{n\in \mathbb{Z}} \delta (t-nT)\Big[(\pi-\theta_e) S^{x}_0 + (\pi-\theta_n) \I_t^{x}\Big],  \label{eq:Hp}
 \end{align}
 where $\I_t^{\pm}= \I_t^{x} \pm i\I_t^{y} $, $S_0^{\pm}= S_0^{x} \pm iS_0^{y} $, $T =2\pi/\omega$ is the time span between pulses, and $\theta_e$ and $\theta_n$ are errors in the pulse rotation angles that shift them away from the ideal value of $\pi$. In color center, rare-earth ion, or quantum dot experiments, the pulses could be implemented using a microwave or optical drive on the electron and a separate RF drive on the nuclei~\cite{morley2013quantum,bradley2019ten,goldman2020optical,wood2021quantum,debroux2021quantum}. Although the use of separate drives allows for the possibility that $\theta_e\ne\theta_n$, we will set $\theta_e=\theta_n=\theta$ for simplicity unless otherwise stated. The fact that the Hamiltonian is time-periodic with period $T$, $H(t+T)=H(t)$, means that we can study the dynamics using Floquet theory. The Floquet operator $U_F(\theta)=U(T,\theta)$, which is the evolution operator that evolves the system over one drive period, can be expressed in terms of the Floquet Hamiltonian, $H_F(\theta)$:
 \begin{align} \label{eq:UT}
 U_F(\theta)=e^{-iH_F(\theta) T}=e^{-i(\pi-\theta)\big( S^{x}_0 +\I_t^{x}\big)}e^{-iH_0 T}.
 \end{align}
 We use the abbreviated notation  $U_F=U_F(\theta=0)$ and $ H_F=H_F(\theta=0)$ to denote these quantities in the absence of pulse errors throughout this paper.

In the context of stroboscopic dynamics, where we focus only on the evolution at a discrete set of times corresponding to the moments immediately after each pulse, if an operator commutes with the Floquet Hamiltonian, then it corresponds to a symmetry of the system. We note that  $\big[H_p, S_0^z+\I_t^z\big]\neq 0\Rightarrow \big[S_0^z+\I_t^z, H_F(\theta)\big]\neq 0$, and therefore $S_0^z+\I_t^z$ is no longer conserved when the system is driven. On the other hand,  $[\I_t^2,\I_t^\mu]=0$ implies that  $[\I_t^2,H_0]=0$ and  $[\I_t^2,H_{\text{p}}]=0$, which implies that
 \begin{equation}
 [H_F(\theta), \I_t^2]=0  \text{ and } [U_F(\theta), \I_t^2]=0, 
 \end{equation}
 and eigenvalues of $\I_t^2$ remain conserved for the stroboscopic dynamics.
The full Hilbert space can be written as $\mathcal{H}=\mathcal{H}_C\otimes\mathcal{H}_S$, where $\mathcal{H}_C$ and $\mathcal{H}_S $ represent the central and satellite subspaces, respectively. Due to the $\I_t^2$ symmetry, we can decompose the satellite part of the Hilbert space into a direct sum of disconnected symmetry sectors, each labeled by a quantum number $j$ of $\I_t^2$, i.e., $\mathcal{H}_S= (\oplus_{j} \mathcal{H}_j)$ (gray boxes in Fig.~\ref{fig:Fig1_Schematic_Diag}(c) show the block diagonalization of the Floquet operator due to this decomposition). For $N$ spin$-\frac{1}{2}$ satellite spins,  the largest total satellite spin subspace is $(N+1)$-dimensional. Thus the full dimension of the largest symmetry subspace when we include a spin$-\frac{1}{2}$ central spin is $2(N+1)$.  We utilize the eigenstates of $\I_t^z$  to form bases for each symmetry sector $j$: $\ket{\psi}=\ket{jm}\ket{\sigma}\equiv \ket{m_j\sigma}$, where $m_j\in \{-j, -j+1,...,j-1, j\}$ is an eigenvalue of $\I_t^z$, and $\sigma=\uparrow \text{or} \downarrow$ labels states of the central spin.

\subsection{Floquet-Krylov subspaces and Hilbert space fragmentation} \label{sec:Hilbert_Frag}
We now examine the emergence of HSF in the stroboscopic dynamics of our periodically driven central-spin model using the concept of Floquet-Krylov subspaces.

Before we define Floquet-Krylov subspaces and compute them for our driven central-spin problem, we first review the ordinary Krylov subspaces that are used to study the continuous-time dynamics generated by a time-independent Hamiltonian. For example, in the absence of driving, our central-spin Hamiltonian $H_0$ from Eq.~(\ref{eq:H0}) generates the evolution operator $e^{-iH_0t}=\sum_{n=0}^{\infty}\frac{(-iH_0t)^n}{n!}$. For a given initial state $\ket{\phi}$, we can define the associated Krylov subspace according to
\begin{align} \label{eq:Krylov_Sub_H0}
     \mathcal{K} (H_0,\ket{\phi}) &= \text{Span of } \Big\{ \ket{\phi},H_0\ket{\phi},H_0^2\ket{\phi},..., \nonumber \\
     &\hspace{23mm}  H_0^n\ket{\phi}, H_0^{n+1}\ket{\phi},...\Big\},
 \end{align}
 where $n$ represents a positive integer. Note that we can equivalently define the subspace by applying powers of the evolution operator $e^{-iH_0t}$ to $\ket{\phi}$. The resulting subspace is identical; we review the proof of this in Appendix~\ref{sec:Appen_Floq_Krylov}.
To study the Krylov subspace of an initial state $\ket{\phi}=\ket{m_j \sigma}$, we split the Hamiltonian into two parts, $H_0= H_{xy}+ H_{z}$, such that $H_{xy}= A_{xy}( \I_t^{+}S_{0}^{-}+\I_t^{-}S_{0}^{+}) $ and
$H_{z}= A_{z}( \I_t^{z}S_{0}^z) + B_{z} S_{0}^z$. 
Because the state $\ket{m_j \sigma}$ is an eigenstate of  $H_{z}$,  repeatedly applying $H_{z}$ to this state only changes it up to a rescaling.  On the other hand, when $H_{xy}$ acts on the state repeatedly, we obtain a two-dimensional (2D) subspace because $(S^{\pm}_0)^2 \ket{\sigma}=0$, where $\ket{\sigma}\in \{\ket{\uparrow},\ket{\downarrow}\}$, and  $S^{+}_0 (S^{-}_0)$ couples with $\I_t^{-} (\I_t^{+})$. Thus, the Krylov subspace  $\mathcal{K} (H_0,\ket{m_j \uparrow})$ is spanned by $\{\ket{m_j \uparrow}, \ket{(m_j+1) \downarrow}\}$ and  $\mathcal{K} (H_0,\ket{m_j \downarrow})$ by $\{\ket{m_j \downarrow}, \ket{(m_j-1) \uparrow}\}$, except $\ket{j\uparrow}$ and $\ket{-j\downarrow}$ as they are eigenstates of the Hamiltonian $H_0$.
Another way to understand the dimension of the Krylov subspace is through the conservation of $\I_t^z +S_0^z$: since $S_0^z$ can take only two values, $\I_t^z$ can also take only two values such that $\I_t^z +S_0^z$ is constant.   

We can always decompose a Hilbert space in terms of the Krylov sectors as $\mathcal{H}=\oplus_{i} \mathcal{K}_{i}$, where $\mathcal{K}_{i}= \mathcal{K}(H_0,\ket{\Psi_i})$ such that each $\mathcal{K}_{i}$ is distinct from the rest, where the initial state $\ket{\Psi_i}$ is a basis state of the full Hilbert space. When there is an exponential number of distinct Krylov sectors corresponding to a set of simple (product or experimentally accessible) states, then we say that the system exhibits HSF \cite{Sala2020Hilbert_Frag,Khemani_2020_Hilbert_Shatter}.
In the case of our undriven central-spin model, we can choose the initial states  $\ket{\Psi_{mj}}=\ket{m_j\sigma}$ for all values of the quantum numbers $j$, $m$, and $\sigma$. 
Since the maximum dimension of a Krylov sector is two,
there is an exponential number ($2^{N-1}$) of distinct Krylov sectors that are dynamically disconnected and span the Hilbert space of $N$ spins (including both satellite and central spins). Thus, the undriven, homogeneous central-spin model of XXZ-type exhibits HSF.

In this work, however, we are more interested in the periodically driven central-spin model, which includes both the undriven part $H_0$, and the driving terms $H_p(t)$ (Eq.~\eqref{eq:Hp}) of the Hamiltonian.  In this case, we can apply the concept of Floquet-Krylov subspaces, which are generated by applying the Floquet operator $U_F$ to some initial state $\ket{\psi}$ repeatedly:
\begin{align} \label{eq:Floq_Krylov}
     \mathcal{K}_F (U_F,\ket{\psi}) &= \text{Span of } \Big\{ \ket{\psi},U_F\ket{\psi},U_F^2\ket{\psi},..., \nonumber \\
     &\hspace{22mm}  U_F^n\ket{\psi}, U_F^{n+1}\ket{\psi},...\Big\}.
 \end{align}
We may naively think that, similar to the case of time-independent Hamiltonians discussed above, the subspace generated by applying powers of the Floquet operator $U_F$ is the same as the Krylov subspace generated by powers of $H_F$, the Floquet Hamiltonian. However, this is not true, because now we are considering the stroboscopic dynamics in which we only keep track of the evolution at integer multiples of the driving period, and therefore the Krylov subspace that corresponds to $U_F$ is a subspace of the Krylov subspace corresponding to $H_F$. (See Appendix~\ref{sec:Appen_Floq_Krylov} for a detailed discussion of this point.) 

We now compute the Floquet-Krylov subspaces of our driven central-spin model, focusing on the case without pulse errors ($\theta=0$). For Ising-type interactions ($A_{xy}=0$), we obtain a 2D Floquet-Krylov subspace ($\mathcal{K}_{F}^Z$) for the initial state $\ket{m_j \sigma}$:
 \begin{align} 
     \mathcal{K}_{F}^Z (U_F,\ket{m_j \sigma}) &= \text{Span of } \Big\{ \ket{m_j \sigma},\ket{-m_j \bar{\sigma}} \Big\}.
\end{align}
Next consider non-Ising interactions (XX-, XXX-, and XXZ-type). When the initial state is $\ket{j\uparrow}$ or $\ket{-j\downarrow}$, the Floquet-Krylov subspace is again 2D and spanned by these two states.
 For other initial states $\ket{m_j \uparrow}$ or $\ket{m_j\downarrow}$ where $m_j\ne j,-j$, the Floquet-Krylov subspace is 4D (see Appendix~\ref{sec:Appen_Floq_Krylov}):
 \begin{align} 
     \mathcal{K}_F (U_F,\ket{m_j \uparrow}) &= \text{Span of } \Big\{ \ket{m_j \uparrow},\ket{(m_j+1) \downarrow}, \nonumber \\
     &\hspace{10mm} \ket{-m_j \downarrow}, \ket{-(m_j+1) \uparrow}\Big\}, \label{eq:Floq_Krylov1} \\
     \mathcal{K}_F (U_F,\ket{m_j \downarrow}) &= \text{Span of } \Big\{\ket{m_j-1 \uparrow},\ket{m_j \downarrow}, \nonumber \\
     &\hspace{10mm}  \ket{-(m_j-1) \downarrow}, \ket{-m_j \uparrow}\Big\}. \label{eq:Floq_Krylov2}
 \end{align} 
This fragmentation of $U_F$ is depicted in Fig.~\ref{fig:Fig1_Schematic_Diag}(c) (blue boxes).
The evolution in each 4D Floquet-Krylov subspace consists of oscillations between two disjointed 2D subspaces. For example, the initial state $\ket{m_j \uparrow}$ first evolves in the 2D subspace spanned by $\ket{m_j \uparrow}$ and $\ket{(m_j+1) \downarrow}$. The first $\pi$ pulse then maps this into the separate 2D subspace spanned by $\ket{-m_j \downarrow}$ and $\ket{-(m_j+1) \uparrow}$, after which the spin-spin interactions rotate the state in this subspace. Each subsequent $\pi$ pulse flips the state from one 2D subspace to the other, such that each subspace is visited every two driving periods. This $2T$ oscillatory behavior breaks the discrete time-translation symmetry of the time-periodic Hamiltonian. To classify this phenomenon more formally, we introduce the notion of subspace time-crystallinity for general time-periodic Hamiltonians in Sec.~\ref{sec:Subspace_TC_Definition}.

We see that even if we increase the number of satellite spins $N$, the Floquet-Krylov subspace remains 4D, while the dimension of the largest symmetry sector corresponding to $\I_t^2$ increases linearly as $N+1$, and the full Hilbert space dimension increases exponentially as $2^N$.
Thus, the ratio of the dimension of the largest stroboscopic dynamical subspace to the dimension of the full Hilbert space (or symmetry sector) goes to zero in the large $N$ limit. This type of fragmentation is the Floquet version of strong HSF \cite{Sala2020Hilbert_Frag} and leads to a non-ergodic behavior of the system. Note that this fragmentation holds for all values of the spin-spin couplings and Zeeman field strengths in the absence of pulse errors. Next, we show that the $2T$ oscillatory behavior is stable against small perturbations $\theta$ to the pulse angle, indicating the existence of time-crystalline behavior in the system without any random coupling disorder.

\begin{figure}[t] 
   \includegraphics[width=3.5in]{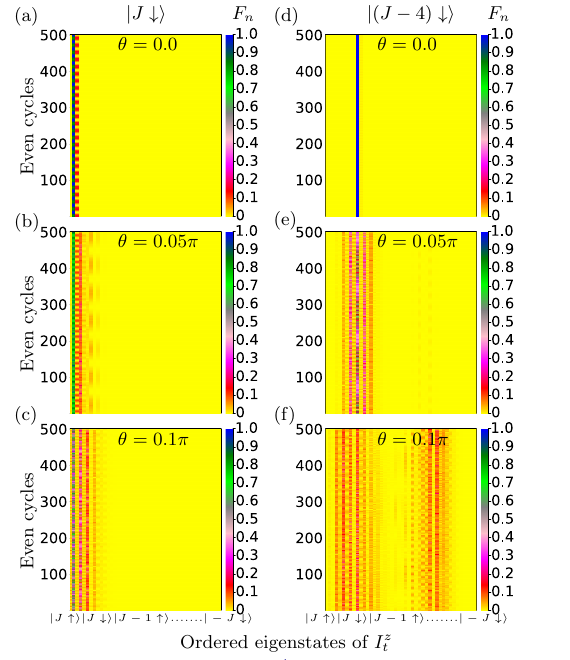}
  \caption{Expansion of Floquet-Krylov subspaces in a Heisenberg central-spin model with $N=21$ satellite spins. The overlaps (color bar) of the state $U_F^n(\theta)\ket{\psi}$ with each of the $2(2J+1)$ basis states in the symmetry subspace with total satellite spin $J=N/2$ (see Eq.~\eqref{eq:overlap}) are shown for two different initial states $\ket{\psi}$ and for 500 different values of $n$ given by $n=10^5-2k$, where $k=0,...,499$. The basis states $\{\ket{m_j\sigma}\}$ are ordered from left to right along the horizontal axes in each panel in the order of decreasing values of $m_j$, alternating between $\sigma=\uparrow$ and $\downarrow$ for the central spin. (a,b,c) The overlaps for $\ket{\psi}=\ket{J\downarrow}$ for three different values of the pulse error $\theta$. (d,e,f) The overlaps for $\ket{\psi}=\ket{(J-4)\downarrow}$. The parameter values are $A_z=A_{xy}=1.3$ MHz, $B_z=100$ MHz, $\omega=1$ MHz.}
  \label{fig:Krylov_Jdown}
\end{figure}

To study the effect of small pulse errors on the Floquet-Krylov oscillations, we first define $U_F(\theta)= U_\theta U_F $,  where $U_\theta = e^{i\frac{\theta}{2}P}$, with $P= \I_t^{+}+\I_t^{-}+\sigma_0^x$. After $n$ Floquet cycles, we can expand the perturbed Floquet operator $U(nT,\theta)$ in powers of $\theta$:
\begin{align}
    &U(nT,\theta) = U_F^n +\frac{i\theta}{2}  \sum_{\ell=0}^{n-1} U_F^j P U_F^{n-\ell}  -\frac{\theta^2}{8}\Big(\sum_{\ell=0}^{n-1}  U_F^\ell P^{2}U_F^{n-\ell}\nonumber\\
    &\hspace{10mm}+ 2\sum_{\ell=0}^{n-2} \sum_{k=1}^{n-\ell-1} U_F^\ell P U_F^k PU_F^{n-\ell-k}\Big) +O(\theta^3).
\end{align}
Notice that applying $P$ on $\ket{m_j\sigma}$ produces a linear combination of the three states $\ket{(m_j\pm 1)\sigma}$, $\ket{m_j\bar{\sigma}}$. In general, each contribution to the perturbed evolution at the $r$th order involves $r$ copies of $P$ distributed among the $n$ copies of $U_F$, and it can thus couple the initial state to up to $\sim3^r$ additional states in the symmetry sector. Thus, for larger values of $\theta$ such that these higher-order contributions dominate, the initial state quickly evolves into a state that has significant overlap with all basis states of the symmetry sector, and the HSF is no longer evident. However, for smaller $\theta$,  the lower orders of the above expansion dominate, and their limited connectivity means that it can take much longer for the initial state to spread throughout the symmetry sector, depending on the type of spin-spin interactions and on the values of the couplings and the Zeeman field. 

To quantify the degree of spreading in Hilbert space, we numerically calculate the extent to which the Floquet-Krylov subspaces expand after a large number of Floquet periods due to the presence of a small pulse error $\theta$. In particular, we calculate the overlap of the state after $n$ Floquet periods (starting from initial state $\ket{\psi}$) with each basis state $\ket{m_j\sigma}$: 
 \begin{equation}\label{eq:overlap}
     F_n(\ket{m_j\sigma},\ket{\psi}) = |\sandwich{m_j\sigma}{U_F^n(\theta)}{\psi}|^2.
 \end{equation}
 Fig.~\ref{fig:Krylov_Jdown} shows all of these overlaps for 500 (large) values of $n$ and for two different initial states, $\ket{\psi} =\ket{J\downarrow}$ and $\ket{(J-4)\downarrow}$, for a central-spin model with Heisenberg interactions ($A_{xy}=A_z$) between the central spin and $N=21$ satellite spins. Here, we consider the largest symmetry sector with total satellite spin $J=N/2$. We see that in the absence of pulse errors, $\theta=0$, the overlaps for both initial states have a high overlap with themselves and a negligible overlap with the other basis states. As we increase the pulse error ($\theta=0.05\pi,0.1\pi$), the overlaps remain highly concentrated for the initial state $\ket{J\downarrow}$ (see Fig.~\ref{fig:Krylov_Jdown}(b,c)), but become more delocalized for the initial state $\ket{(J-4)\downarrow}$ (see Fig.~\ref{fig:Krylov_Jdown}(e,f)). This shows that the HSF is less ergodic and more robust to pulse errors when the initial state is more polarized. In Appendix~\ref{sec:Appen_Floq_Krylov}, we also show how the dimensions of the Floquet-Krylov subspaces grow over time by computing them across a large range of Floquet cycles. There, it is evident that the growth for the state $\ket{J\downarrow}$ is much more subdued compared to that for the state $\ket{(J-4)\downarrow}$.
 
The four states $\{ \ket{\pm J\sigma}\}$ are product states, $\ket{J}\equiv\ket{\uparrow_1\uparrow_2...\uparrow_N}$ and $\ket{-J}\equiv\ket{\downarrow_1\downarrow_2...\downarrow_N}$, unlike the other basis states $\{ \ket{(J-p)\sigma}\}$ with $p\in \mathbb{N}$ and $p<2J$. In what follows, we will mostly focus on these states, both because they exhibit the greatest degree of non-ergodicity, and because the ease in preparing the initial state experimentally is an important consideration for exploring this non-ergodicity in the laboratory. We discuss possible experimental implementations in more detail later on in Sec.~\ref{sec:Dynamical_probe_and_Experiment}.

Although not shown here, results similar to Fig.~\ref{fig:Krylov_Jdown} also occur for other types of interactions. In Appendix~\ref{sec:Appen_Exact_Ising_XX}, we analytically compute the unperturbed stroboscopic evolution of a state $\ket{m\sigma}$ for Ising interactions with a central-spin Zeeman field and for XX interactions without a Zeeman field. We also calculate the perturbed stroboscopic evolution up to ${\cal O}(\theta)$ for even cycles (for Ising) and for two cycles (for XX). These analytical results reveal how the overlaps of the initial state $\ket{m\sigma}$ with the various symmetry sector basis states depend on parameters like couplings, the number of satellite spins, the total satellite spin $z$-projection ($m$), and the Zeeman field (in the Ising case). These results will be used later to shed light on additional numerical findings.
\subsection{Evidence of scar states from entanglement entropy} \label{sec:Scar_Diagonosis_XXX} 
\begin{figure}[t]
   \centering
    \includegraphics[width=3.7in]{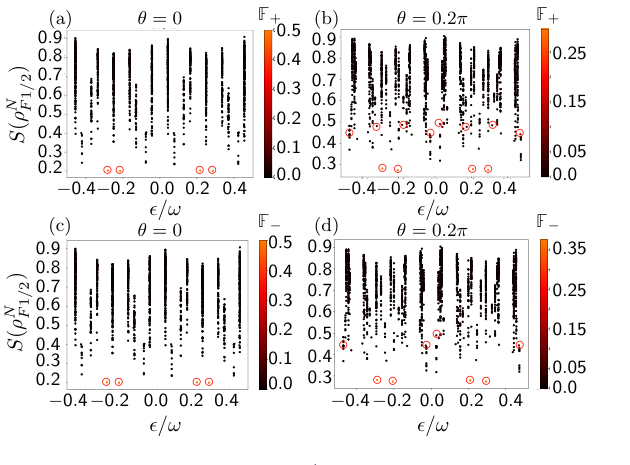}       
  \caption{Bipartite entanglement entropy ($S(\Tr_{N/2}\rho^N_{F}=\rho^N_{F1/2})$) of the satellite part of each Floquet eigenstate ($\rho^N_F$) and its overlap with the fully polarized satellite states $\ket{\pm J}$ for a central-spin model with $N=10=2J$ satellite spins and Heisenberg interactions. Each Floquet eigenstate is labeled by its quasienergy ($\epsilon/\omega$, horizontal axis). The vertical axis in each panel shows the bipartite entanglement entropy corresponding to an equipartition of the satellite part of each Floquet eigenstate. The color bar indicates the overlap $\mathbb{F}_\pm$ of each eigenstate with the state $\ket{\pm J}$ (see Eq.~\eqref{eq:FloquetOverlaps}). Panels (a,b) show $\mathbb{F}_+$, while (b,c) show $\mathbb{F}_-$. Results without (a,c) and with (b,d) pulse errors are shown. The circled red dots are points of maximal $\mathbb{F}_\pm$.
The parameters are $A_{xy}=A_z=\sqrt{2} $ MHz, $B_z$= 100 MHz, $\omega=1$ MHz.}
   \label{fig:Floq_Scar}
\end{figure}
Here, we demonstrate that the fully polarized satellite states ($\ket{\pm J}$) are related to Floquet scar states. To do this, we need to show that the fully polarized states have a substantial overlap with Floquet eigenstates with atypically low bipartite entanglement entropy. We proceed by first sorting the eigenstates of $H_F$ in increasing order of quasienergy ($\epsilon/\omega \in (-0.5.,0.5]$). For each Floquet eigenstate $\ket{\psi}$, we trace out the central spin to obtain the reduced density matrix of all the satellite spins:  $\rho_{F}^N= \Tr_C (\ket{\psi}\bra{\psi})$. We then split the satellite spins into two equal sets of size $N/2$ (see Fig.~\ref{fig:Fig1_Schematic_Diag}(d)) and compute the bipartite entanglement entropy $S(\Tr_{N/2}\rho^N_{F})$ as well as the overlap with the fully polarized states $\ket{\pm J}$:
 \begin{equation}\label{eq:FloquetOverlaps}
     \mathbb{F}(\ket{\pm J},\rho^N_F)= \sandwich{\pm J}{\rho^N_F}{\pm J} \equiv \mathbb{F}_{\pm}.
 \end{equation}

Results for the central-spin model with Heisenberg interactions are shown in Fig.~\ref{fig:Floq_Scar}. In particular, the figure shows the entanglement entropy of each Floquet eigenstate and its overlap with both fully polarized satellite states with ($\theta=0.2\pi$) and without ($\theta=0$) pulse errors. We observe that the orange-colored (circled) dots (eigenstates) possess atypically low bipartite entanglement entropy and have high overlap with the satellite states $\ket{\pm J}$.  Therefore, for both values of $\theta$, the satellite states $\ket{\pm J}$ are closely associated to scar states (Floquet eigenstates with atypically low bipartite entanglement entropy) of the system. Additionally, in Appendix~\ref{sec:Appen_Floq_Scar_XXZ}, we show the entanglement entropy of Floquet eigenstates of the driven XXZ central-spin system. There too, fully polarized satellite states exhibit the characteristics of scar states.

\section{Scar time-crystals} \label{sec:Scar_TC}

Next, we show that periodically driven central-spin models can exhibit (subspace) scar time-crystalline behavior in certain parameter regimes. 
In this section, we first define subspace time crystals using the concept of Floquet Krylov subspaces for a general time-periodic Hamiltonian and discuss signatures of time-crystallinity in our model.
We then show that only a few initial states that exhibit stable HSF and have high overlap with Floquet scar states exhibit a robust period doubling of the total satellite spin magnetization. Therefore the system shows scar time-crystalline behavior. 
 We then consider the four initial states ($\{\ket{\pm J\sigma}\}$) and investigate the time crystallinity with respect to coupling strength, Zeeman energy of the central spin, and pulse error. 
\subsection{Subspace time crystal} \label{sec:Subspace_TC_Definition}
We consider a time-periodic Hamiltonian $H(t+T)=H(t)$ in a Hilbert space $\mathcal{H}$. We decompose this Hilbert space in terms of Floquet Krylov subspaces as $\mathcal{H}=\oplus_{i} \mathcal{K}_{Fi}$, where $\mathcal{K}_{Fi}=\mathcal{K}_{F}(U_F,\ket{\psi_i})$ is defined in Eq.~(\ref{eq:Floq_Krylov}).
Each $\mathcal{K}_{Fi}$ consists of disjointed subspaces such that                   $\mathcal{K}_F= \oplus_{q=1}^{r}\mathcal{K}_F^q$, where $r\in\mathbb{N}$. These disjointed subspaces are obtained from the spans of states reached after integer multiples of the drive period. For an experimentally accessible quantum state (used in defining the Floquet Krylov subspace) and $n\in \mathbb{N}$, if we find that
\begin{align}
    &U_F^{n}\ket{\psi}\in \mathcal{K}_F^q, \text{ but } U_F^{n+r_0}\ket{\psi}\notin \mathcal{K}_F^q \ \forall \ r_0<r \in\mathbb{N}, \text{ and} \nonumber\\
    &\hspace{25mm} U_F^{n+r}\ket{\psi}=\ket{\psi'}\in \mathcal{K}_F^q,
\end{align}
 then the system periodically returns to the same subspace with period $rT$ and thus breaks the discrete time translation symmetry. When this breaking remains stable under small local perturbations, we say the system exhibits subspace time-crystallinity. If all subspaces show the robustness against the perturbations then we have ``full (MBL type) subspace time-crystallinity''. But if the robustness exists only for a few states, then we have ``subspace scar time-crystallinity''. We note that for the existence of a subspace TC, some Floquet-Krylov subspaces must contain at least two disjointed subspaces ($r>1$). Additionally, we can recast a conventional (MBL) TC in terms of two dimensional Floquet-Krylov subspaces containing two 1D disjointed subspaces.
 
To diagnose a subspace TC, we can calculate the probability (fidelity) to return to the subspace as  $\mathbb{F}^n_\Sigma= \sum_{i}|\sandwich{\phi_i}{U_F^n}{\psi}|^2$, where $\{\ket{\phi_i}\}$ are orthonormal bases of the disjointed subspace containing $\ket{\psi}$. When the system periodically returns to the initial subspace after $r$ drive periods, then we obtain  $\mathbb{F}^n_\Sigma= \delta_{n (jr)}$, $j\in\mathbb{N}$. Other diagnostic probes such as an expectation value of an observable in time may not be useful as the evolved state may be a superposition of eigenstates of the observable. However, in some cases (in our model too), some observable (net magnetization) works well as the dimension of the disjointed subspace is small (two-dimensional in our model).
 
We can further outline a general framework to identify subspace time-crystallinity whenever there is Hilbert space fragmentation for a time-independent Hamiltonian. Consider a time-independent Hamiltonian $H_0$ that exhibits HSF, so that $\mathcal{H}=\oplus_i \mathcal{K}_i$, where $\mathcal{K}_i=\mathcal{K}(H_0, \ket{\psi_i})$ defined in Eq.~(\ref{eq:Krylov_Sub_H0}). We choose two non-ergodic subspaces (say $\mathcal{K}_1, \mathcal{K}_2$) such that we can connect them using some local unitary operator ($U_L$), i.e., if a state  $\ket{\psi}\in \mathcal{K}_1$ then $U_L\ket{\psi} \in \mathcal{K}_2$ and vice-versa.  Thus, for Floquet operator $U_F=U_Le^{-iH_0T}$, we obtain a subspace TC ($\mathcal{K}_F(U_F,\ket{\psi})=\mathcal{K}_1\oplus \mathcal{K}_2 $)  if it also shows robustness against small local perturbations.

Now we focus on our central spin model.
We define the (subspace) scar time-crystal as a non-equilibrium phase that shows a  robust subharmonic response only for a few initial states associated with the weak breaking of ergodicity.
To diagnose the time-crystalline behavior, we need to calculate the average satellite spin magnetization at integer multiples of the driving period. For an  initial state $\ket{\Psi(0)}$, we have 
\begin{equation}
  \braket{\I_{t}^z(nT)} =\frac{1}{N}\sandwich{\Psi(nT)}{\I_t^z}{\Psi(nT)},   
\end{equation}
where  $\ket{\Psi(nT)} =U(nT)\ket{\Psi(0)}$, $N$ is the number of satellite spins, and $\I_t^z=\sum _{p=1}^N \I_p^z$ as before. When $ \braket{\I_t^z((n+1)T)}\neq \braket{\I_t^z(nT) }$ for large $n$, then the average magnetization explicitly breaks the discrete time-translation symmetry of the system. This is a signature of the time-crystalline behavior of the system. We note that for a subspace TC, ideally we should examine the return probability ($\mathbb{F}^n_\Sigma$) as discussed above. However, because the disjointed subspaces (Krylov subspaces of $H_0$) are maximally two-dimensional and eigenstates of $\I_t^z$, the net satellite magnetization works well. Henceforth in this text,  we use the term ``scar TC'' to refer to what is in fact a subspace scar TC for convenience. We will discuss the nature of time-crystallinity corresponding to different interactions and initial states in Sec.~\ref{subsec:Nature of TC}.

\begin{figure}[t]
   \centering
   \includegraphics[width=3.4in]{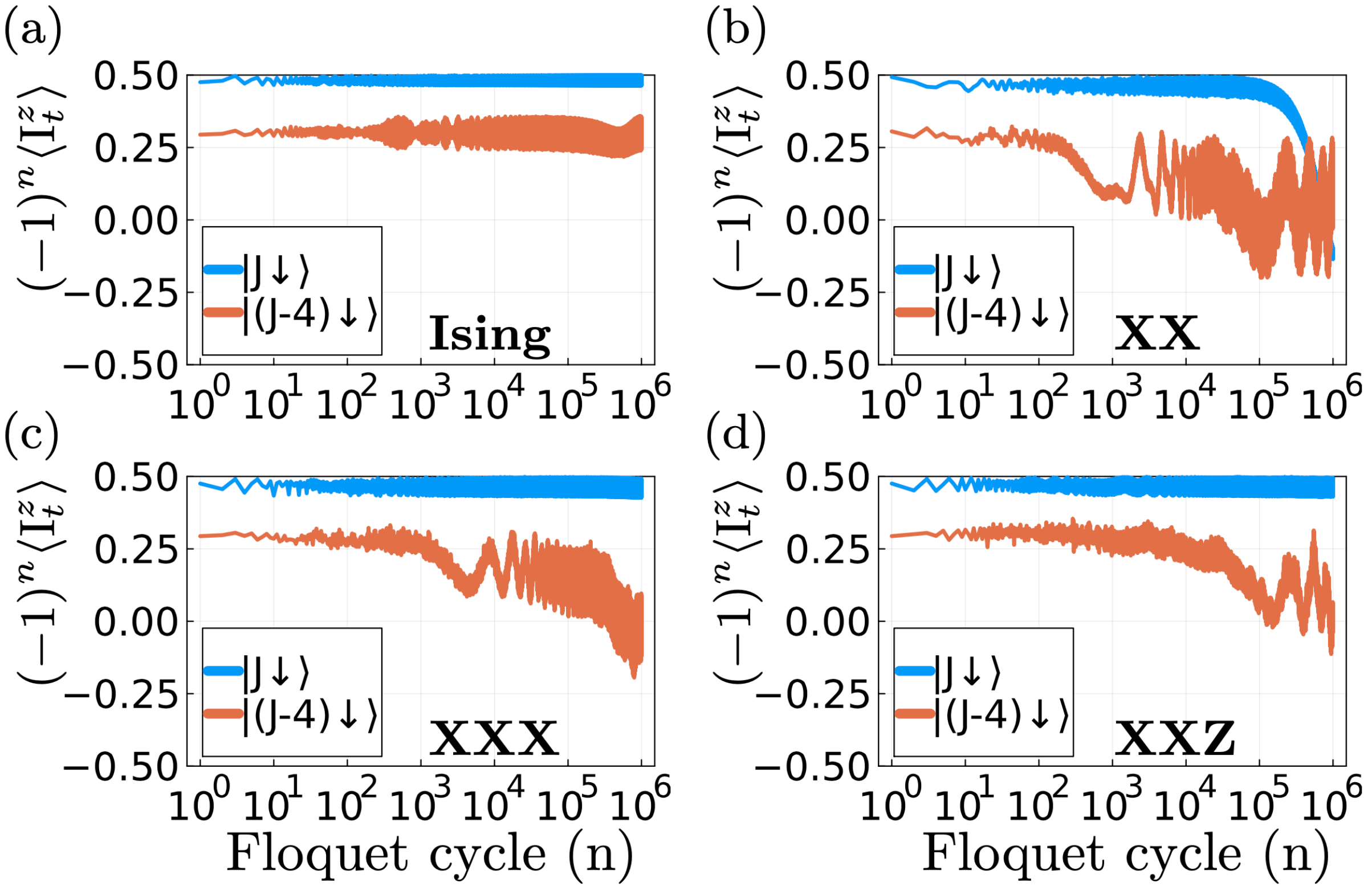}
  \caption{ Time-staggered satellite spin magnetization for four different types of interactions and two different initial states $\ket{J\downarrow}, \ket{(J-4)\downarrow}$, and for $N=21$ satellite spins. (a) Results for Ising interactions with $A_{xy}=0$, $A_{z}=1.3$ MHz, $B_z=100$ MHz, $\theta=0.1\pi$. (b) Results for XX interactions with $A_{xy}=1.3$, $A_z=0$, $B_z=0$, $\theta=0.05\pi$. (c) Results for Heisenberg interactions with $A_z=A_{xy}=1.3$ MHz, $B_z=100$ MHz, $\theta=0.1\pi$. (d) Results for XXZ interactions with $A_{xy}=1.3$ MHz, $A_z=3.3$ MHz,  $B_z=100$ MHz,  $\theta=0.1\pi$. $\omega=1$ MHz in all cases.}
   \label{fig:Fig_Iz_Jdown_Jminus4}
\end{figure}
\subsection{Scar time-crystallinity in Ising, XX, XXX, and XXZ central-spin models}
We consider Ising, XX, Heisenberg (XXX), and XXZ types of interactions and investigate the time-crystalline behavior that emerges for different initial states. We calculate the expectation value of the time-staggered average satellite spin magnetization ($(-1)^n\braket{I^z_t (nT)}$); the degree to which this quantity remains constant serves as an indicator for period doubling and thus the breaking of discrete time-translation symmetry.

Figure~\ref{fig:Fig_Iz_Jdown_Jminus4} shows the staggered magnetization as a function of time (in terms of Floquet periods) for four different types of interactions and for initial states $\ket{J\downarrow}$ and $\ket{(J-4)\downarrow}$, with $N=2J=21$ satellite spins.  For Ising interactions, both initial states show a robust time-crystalline behavior that survives for at least a million Floquet periods [Fig.~\ref{fig:Fig_Iz_Jdown_Jminus4}(a)]. On the other hand, for Heisenberg (XXX) and XXZ-type interactions, time-crystallinity is stable for $\ket{J\downarrow}$ but not for $\ket{(J-4)\downarrow}$ on long time scales [Fig.~\ref{fig:Fig_Iz_Jdown_Jminus4}(c,d)]. For XX interactions and in the absence of a Zeeman field, time-crystalline behavior persists for a much longer time for $\ket{J\downarrow}$ compared to $\ket{(J-4)\downarrow}$ [Fig.~\ref{fig:Fig_Iz_Jdown_Jminus4}(b)], although it does eventually decay after $10^5$ periods. (We set the Zeeman field to zero in this case because it provides more stable time-crystallinity, as we show below.) Thus, an eigenstate of $\I_t^z$ with extreme eigenvalues (highly polarized states) show more robust time-crystalline behavior compared to less polarized eigenstates regardless of the type of interaction. 
We can understand this phenomenon based on the stability of HSF discussed in section \ref{sec:Hilbert_Frag}.  Furthermore, the discussion in Sections \ref{sec:Hilbert_Frag} and \ref{sec:Scar_Diagonosis_XXX} highlights the weak breaking of ergodicity, suggesting that the time crystalline behavior for a fully polarized nuclear state, observed in Fig.~\ref{fig:Fig_Iz_Jdown_Jminus4}, will persist for much longer than what is shown.

Recently, a conventional (MBL-type) time-crystalline behavior was shown to occur in Heisenberg and XXZ-type central-spin models with random coupling disorder for all initial product states \cite{Rafail_PRB_2023}. In that work, the ergodicity breaking was likely related to MBL since random coupling disorder was necessary to achieve stable non-ergodicity. Additionally, the emergence of time-crystallinity required either a large Zeeman energy difference between the central and satellite spins or the application of additional pulses on the central spin every Floquet period. In contrast, here the time-crystalline behavior is due to stable HSF for fully polarized satellite states that have high overlap with Floquet scar states.  Therefore, here time-crystallinity does not require random coupling disorder but only a large Zeeman energy difference between central and satellite spins, and it arises for only a few initial states. Interestingly, we see in Fig.~\ref{fig:Fig_Iz_Jdown_Jminus4}(b) that in the case of the XX central-spin model, a Zeeman energy difference is not necessary for time-crystallinity for the initial state $\ket{J\downarrow}$. We further examine the behavior of the XX model under random coupling disorder (Gaussian distribution) for $B_z=0$ in Appendix~\ref{sec:Appen_XX_random_bond}, where we find that as we increase the disorder in the coupling terms, the stability of the time-crystallinity decreases. Thus, the XX central-spin model does not show the conventional (MBL type) time-crystalline behavior, but does exhibit scar time-crystallinity. We further discuss the nature of scar time-crystallinity in Sec.~\ref{subsec:Nature of TC}. In the next subsection, we will see that for more general XXZ interactions, we can also observe scar time-crystallinity without a Zeeman energy difference. 
\subsection{Phase diagrams}\label{sec:phase_diagm}
In this section, we map out the time-crystalline phase region for periodically driven Heisenberg (XXX), XXZ, XX, and Ising central-spin models for the initial state $\ket{J\uparrow}$.
We use the time-averaged staggered magnetization,
\begin{equation}\label{eq:time-averaged_magnetization}
    \bbraket{\I_t^z}= \frac{1}{N_{C}}\sum_{n=1}^{N_{C}}  (-1)^{n}\sandwich{\Psi(nT)}{\I_t^z}{\Psi(nT)},
\end{equation}
as an order parameter, where $N_C$ is the total number of Floquet cycles. If  $\bbraket{I_t^z}\approx 0.5$ for large $N_C$ even in the presence of pulse errors, then this indicates robust time-crystalline behavior. This also means that the time average of the magnetization does not converge to one value and therefore cannot be equal to an ensemble average. Naturally, this also illustrates the non-ergodicity of the system when it is initialized appropriately.  

\begin{figure}[t]
   \centering
   \includegraphics[width=3.4in]{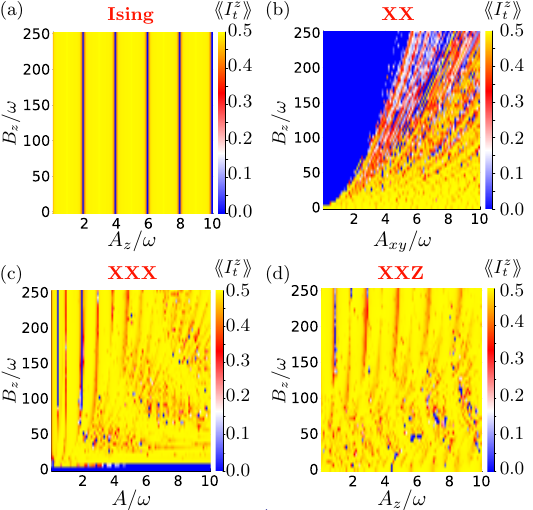}
  \caption{Non-equilibrium phase diagrams: Staggered satellite total spin time-averaged over $N_C=10^4$ Floquet cycles (Eq.~\eqref{eq:time-averaged_magnetization}) as a function of the Zeeman field ($B_z$) and interaction strength ($A_z$ or $A_{xy}$ or $A=A_z=A_{xy}$). Phase diagrams are shown for (a) Ising, (b) XX, (c) Heisenberg, (d) XXZ interactions. The parameters are $\theta=0.03\pi$, $\omega=1$ MHz. In (d) $A_{xy}=4.5$ MHz. The initial state is $\ket{J\uparrow}$ and $N=2J=21$ in all cases.}
   \label{fig:fig5_ScarTC_phase_B_A}
\end{figure}

\begin{figure}[t]
   \centering
   \includegraphics[width=3.4in]{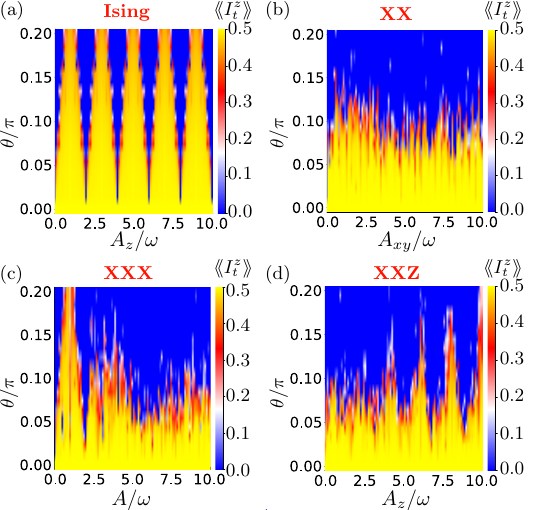}
  \caption{Non-equilibrium phase diagrams: Staggered satellite total spin time-averaged over $N_C=10^4$ Floquet cycles (Eq.~\eqref{eq:time-averaged_magnetization}) as a function of the pulse error ($\theta$) and interaction strength ($A_z$ or $A_{xy}$ or $A=A_z=A_{xy}$). Phase diagrams are shown for (a) Ising, (b) XX, (c) Heisenberg, (d) XXZ interactions. The parameters are $B_z=100$ MHz (in (a,c,d)), $B_z=0$ MHz (in (b)), $\omega=1$ MHz. In (d) $A_{xy}=4.5$ MHz. The initial state is $\ket{J\uparrow}$ and $N=2J=21$ in all cases.}
   \label{fig:fig6_ScarTC_phase_theta_A}
\end{figure}

In Figs.~\ref{fig:fig5_ScarTC_phase_B_A} and \ref{fig:fig6_ScarTC_phase_theta_A}, we show two sets of non-equilibrium phase diagrams. In the first set (Fig.~\ref{fig:fig5_ScarTC_phase_B_A}), we show the time-averaged staggered magnetization ($N_C=10^4$) as a function of the central-spin Zeeman energy and the coupling strength for different types of interactions. Figure~\ref{fig:fig5_ScarTC_phase_B_A}(a) shows the case of an Ising interaction, where a robust time-crystalline behavior persists across a large range of coupling values, except for a certain discrete set of coupling strengths corresponding to when $A_z$ is an even-integer multiple of the driving frequency. A phase diagram for XX interactions is shown in Fig. \ref{fig:fig5_ScarTC_phase_B_A} (b), where it can be seen that the time-crystalline behavior now depends on the relative strength of the Zeeman energy and the coupling term. For $B_z =0$, the behavior arises for all considered values of the interaction strength $A_{xy}$.  As we increase $B_z$, the system thermalizes more quickly for the lower values of $A_{xy}$, leading to $\bbraket{\I_t^z}\approx 0$. 

Figures~\ref{fig:fig5_ScarTC_phase_B_A}(c) and \ref{fig:fig5_ScarTC_phase_B_A}(d) show phase diagrams for isotropic Heisenberg and anisotropic XXZ interactions, respectively. In the case of Heisenberg interactions,  we observe that at $B_z =0$, the system exhibits ergodic behavior, with $\bbraket{\I_t^z}\approx 0 $. However, for large $B_z$, we obtain $\bbraket{\I_t^z}\approx 0.5$ for most coupling strengths $A$. On the other hand, for XXZ interactions, a robust time-crystalline behavior persists all the way down to $B_z=0$, suggesting that the anisotropy somehow compensates for the vanishing Zeeman energy in this case.
 
In Fig. \ref{fig:fig6_ScarTC_phase_theta_A}, we show a different set of phase diagrams in which we consider the dependence of the staggered magnetization on the pulse error ($\theta$) instead of the Zeeman field. Overall, we see that Ising interactions are much more robust against the pulse error [Fig.~\ref{fig:fig6_ScarTC_phase_theta_A} (a)] especially when $A_z/\omega$ is an odd integer. This is because, for these values of $A_z/\omega$, the first-order perturbative term vanishes in the evolved state and, hence, in $\sandwich{\psi(2pT,\theta)}{\I_t^z}{\psi(2pT,\theta)}$ (see Appendix~\ref{sec:Appen_Ising_Exact_dynamics} for details). 
While  Heisenberg interactions show a similar robustness only for smaller values of the coupling strength [Fig.~\ref{fig:fig6_ScarTC_phase_theta_A}(c)]. In the case of XX interactions without Zeeman field [Fig.~\ref{fig:fig6_ScarTC_phase_theta_A} (b)], the system shows ergodic behavior for all considered values of $A_{xy}$ when $\theta >0.1\pi$; the narrow phase regions that extend to high values of $\theta$ in the case of Ising or Heisenberg interactions are no longer evident here. Anisotropic XXZ interactions [Fig.~\ref{fig:fig6_ScarTC_phase_theta_A} (d)], on the other hand, produce phase regions that resemble those of the XX model for lower values of $A_z$, while they become more like those of the Ising model for larger values of $A_z$. This is as expected since the model interpolates between the Ising and XX models as $A_z$ is swept from zero up to values large compared to $A_{xy}$. 

Finally, we would like to emphasize that, in the Floquet operator, the period ($T=2\pi/\omega$) always appears together with coupling terms ($A_{z},A_{xy}$) or the Zeeman field ($B_z$). This means that we can tune in and out of the scar DTC phase by adjusting $T$.

\begin{table*}[t]
    \centering
    \caption{Summary of when MBL and scar DTCs occur in central-spin systems depending on the type of interaction and other parameters. MBL DTCs require disorder in the couplings, while scar DTCs do not.}
    \label{tab:DTC_table}
    \renewcommand{\arraystretch}{1.0}
    \setlength{\tabcolsep}{10pt}
    \begin{tabular}{p{2.5cm} p{3cm} p{3cm} p{7.5cm}}
    \hline
    \hline
       Interaction Type & MBL-DTC (Random Couplings) & Scar-DTC (Homogeneous Couplings) & Remarks \\
    \hline
       None  & No & No  & DTC does not emerge without interactions.\\
       Ising  & Yes &  Yes & DTC is robust except at certain discrete values of \( A_z/\omega \). Robustness is maximized around odd values of \( A_z/\omega \). \\
       XX  & No & Yes & A robust scar DTC exists at \( B_z = 0 \), but it becomes unstable for large \( B_z \).\\
       Heisenberg (XXX)  & Yes & Yes & Robust MBL and scar DTCs emerge when there is a large Zeeman energy mismatch between the central and satellite spins.\\
       XXZ & Yes & Yes & A robust scar DTC exists. An MBL DTC emerges only when there is a large Zeeman energy mismatch between the central and satellite spins.\\
       \hline
    \hline
    \end{tabular}
    \label{tab:DTC_table}
\end{table*}


\subsection{Scar versus conventional discrete time crystals}
\label{subsec:Nature of TC}

In this section, we elaborate further on the two key features of scar time-crystallinity in homogeneous central-spin models and on how they differ from conventional MBL DTCs~\cite{else2016floquet}: (i) In contrast to conventional DTCs, here only a few initial (scar) states show robust time-crystallinity. Moreover, (ii) in the case of scar time-crystallinity we can have subspaces, and not just individual states, that exhibit $2T$ periodicity, and this subharmonic response can persist in a few of the subspaces despite the presence of pulse errors. We elaborate on these points in the following.

We first discuss point (i). We focus on the case of Ising interactions, where individual states rather than subspaces are preserved. As we saw in Sec.~\ref{sec:Hilbert_Frag}, all states of the form $\ket{m_j\sigma}$ break the discrete-time translation symmetry. However, only highly polarized states like $\ket{J\uparrow}$ and $\ket{-J\downarrow}$ exhibit robustness against pulse errors, and therefore lead to scar time-crystallinity. This is similar to the scar DTC that arises in the driven PXP model \cite{Maskara_PRL_ScarTC2021}, where all product states break the discrete time-translation symmetry, but only the N\'eel states display a robustness against pulse errors. This is also consistent with the observation that the Schr\"odinger cat states $\frac{1}{\sqrt{2}}(\ket{J\uparrow} \pm \ket{-J\downarrow})$ are eigenstates of the unperturbed Floquet operator $U_F$. The fact that these long-range correlated states are not physically accessible implies that if the system starts in a highly polarized sector, it cannot easily equilibrate to an eigenstate of $U_F$. However, unlike in a conventional DTC, not all eigenstates of $U_F$ are Schr\"odinger cat states, in which case they can be reached quickly for some choices of the initial state, leading to the breakdown of a coherent subharmonic response. Conventional DTCs require disorder in the couplings in order to ensure that all eigenstates are long-range correlated and non-degenerate. In contrast, coupling disorder is not necessary in the case of scar DTCs.

Regarding point (ii), we saw in Sec.~\ref{sec:Hilbert_Frag} that for interactions of non-Ising type (XX, XXX, XXZ) and for most initial states of the form $\ket{m_j\sigma}$, there is a 4D Floquet-Krylov subspace that breaks the discrete-time translation symmetry. The eigenstates of $U_F$ within these subspaces again have a Schr\"odinger cat-like form, making them physically inaccessible. However, only subspaces that contain eigenstates of $\I_t^z$ with extreme eigenvalues (highly polarized states) show robustness against pulse errors (see Fig.~\ref{fig:Fig_Iz_Jdown_Jminus4}(b,c,d)). We can think of the conventional DTC as a special case of this ``subspace'' scar DTC in which the dimension of each subspace is one, and where all initial states are scar states.

We conclude this section by summarizing the types of DTC (conventional MBL versus scar) that can arise for each type of central-spin interaction based on the present work and on Refs.~\cite{Rafail_PRB_2023,Pal_PRL_2018}. For Ising interactions, we can observe scar DTCs for most homogeneous longitudinal interaction strengths $A_z$ and MBL DTCs for random couplings $A_z$ \cite{Pal_PRL_2018}. In the presence of only transverse XX interactions, the scar DTC is observed for $B_z=0$, but no MBL DTC occurs, as random couplings adversely affect the time-crystalline behavior (see Appendix ~\ref{sec:Appen_XX_random_bond} for further discussion on this point). In the case of Heisenberg (XXX) interactions, we can realize scar DTCs and MBL DTCs \cite{Rafail_PRB_2023} provided there is a large Zeeman energy mismatch between the central spin and the satellite spins.  Finally, for XXZ interactions, in the regime of a large Zeeman energy mismatch, both types of DTCs can be realized. Additionally, for zero or small Zeeman energy of the central spin, scar DTCs can still be observed. These findings are summarized in Table \ref{tab:DTC_table}.

\section{Experimental implementations}
\label{sec:Dynamical_probe_and_Experiment}

The central-spin models studied in this work naturally arise in a number of physical platforms, including color centers in solids such as nitrogen-vacancy centers in diamond, electron or hole spins confined in semiconductor quantum dots, and rare-earth ions~\cite{Wolfowicz2021,Chatterjee2021,Burkard2023,ruskuc2022nuclear,gangloff2019quantum}. In each case, a central electronic spin couples to several or many nuclear spins via hyperfine interactions. In the context of color centers or rare-earth ions, the electronic wavefunction is spatially localized, and so the electron-nuclear interactions are predominantly dipolar in nature. These anisotropic interactions are similar to the Ising or XXZ-type interactions considered in this work. In contrast, the electronic wavefunction in quantum dots is delocalized, overlapping with $10^4$ to $10^6$ nuclear spins depending on the size of the dot, thus giving rise to a contact hyperfine interaction, which is of Heisenberg type. In the case of nuclei with spin greater than 1/2, a significant amount of strain in the dot can also give rise to anisotropic nuclear quadrupolar interactions, which can dominate the hyperfine interaction in some cases. The $p$-orbital nature of hole spin wavefunctions in quantum dots suppresses the contact interaction, leaving the anisotropic dipolar coupling as the dominant hole-nuclear interaction in this case. Thus, depending on the platform, a variety of central-spin models with isotropic or anisotropic interactions are naturally realized.

A key experimental requirement for realizing DTCs in electron-nuclear central-spin systems is the ability to control both the electronic as well as the nuclear spins. Control of the electronic spin is routinely done in experiments using either microwave or optical pulses. Direct control of nuclear spins has also been demonstrated using separate RF fields, although this approach requires additional experimental overhead~\cite{Munsch2014,bradley2019ten}. Alternatively, one can instead drive nuclear spin rotations indirectly by using carefully designed pulse sequences that are applied to the electronic spin~\cite{Taminiau2012, ruskuc2022nuclear, Takou2023, millington2023approaching,gangloff2019quantum,bourassa2020entanglement}. Nuclear spin control is necessary for both initializing the system in a highly polarized state and for performing the $\pi$ rotations that are needed every Floquet period. Refs.~\cite{Taminiau2012,bradley2019ten,bourassa2020entanglement,ruskuc2022nuclear} demonstrated the initialization and pulsing of a small set of nuclear spins in the vicinity of color centers or rare-earth ions via indirect control through the electronic spin, in some cases with the assistance of an additional RF field. Specifically, in weakly-coupled systems, dynamical decoupling sequences of $\pi$ pulses can be used to coherently interact an electron spin with a single nucleus while decoupling the rest of the system. These sequences enable coherent electron-nuclear dynamics, facilitating full nuclear control across various platforms~\cite{Taminiau2012,bradley2019ten,uysal2023coherent}. Even though many-body dynamics may seem complex, dynamical decoupling can isolate an effective two-body system by decoupling other spins, maintaining the interaction between the electron spin and a single nuclear spin. Ideally, complete manipulation of the nuclear spin could flip the entire magnetization of the spin bath. In the case of quantum dots, where the nuclei constitute a dense spin ensemble, individual addressing of nuclear spins is not possible. However, initialization of the nuclei into highly polarized states is still possible using indirect electron-mediated control~\cite{Jackson2022,millington2023approaching}. Moreover, control of individual nuclear magnon modes in these systems has also been demonstrated~\cite{jackson2021quantum,Jackson2022,denning2019collective}.

Another important issue in realizing scar time-crystallinity in physical central-spin systems is the distribution of electron-nuclear couplings in these systems. In the case of color centers in diamond, for example, spinful $^{13}$C nuclei are randomly distributed throughout the diamond lattice with a natural abundance of 1.1\%. The distance and orientation of each nucleus relative to the color center determines the magnitudes of the hyperfine couplings, and so significant coupling disorder is naturally present in the system. An effective coupling uniformity could be achieved dynamically by adjusting the spacing of pulses applied to the electronic spin in such a way as to select out only a few nuclei with similar couplings to participate in the joint electron-nuclear dynamics. This approach essentially amounts to a dynamical decoupling of all other nuclei from the color center. In the case of rare-earth ion systems such as a $^{171}$Yb$^{3+}$ ion in YVO${}_4$ that couples to nearby $^{51}$V$^{5+}$ nuclei, the 99.75\% natural abundance of the latter ensures that the nuclear spins naturally form sets that uniformly couple to the $^{171}$Yb$^{3+}$ electronic spin~\cite{ruskuc2022nuclear}. The situation is somewhat more complex in quantum dots, where electron-nuclear hyperfine couplings form a dense, Gaussian-like distribution. In this case, one could imagine using dynamical decoupling techniques on the electronic spin to single out annular subsets of nuclei with equal couplings, while decoupling all other nuclear spins.

Overall, there are many promising avenues for the potential experimental realization of scar time-crystalline physics in electron- or hole-nuclear central-spin systems in a variety of platforms, with many of the key capabilities already demonstrated. We also recognize that combining all of these key capabilities may still be challenging, but we are optimistic that this can be achieved in the future.

\section{Conclusion and Outlook} \label{sec:Discussion}
 
We have shown that Hilbert space fragmentation and (subspace) scar time-crystallinity can arise in various types of periodically driven central-spin models with homogeneous interactions. The dynamical formation of two-dimensional or four-dimensional Floquet-Krylov subspaces leads to the spontaneous breaking of discrete time-translation symmetry and gives rise to a subharmonic response in the total satellite spin magnetization. We further showed that this response persists in the presence of pulse errors only for highly polarized initial states---a defining feature of scar time-crystallinity. We mapped out the scar DTC phase regions of Ising, XX, Heisenberg, and XXZ central-spin models by computing the dependence of the time-averaged satellite spin magnetization on the applied magnetic field, the interaction strength, and the pulse error. Unlike conventional MBL-type DTCs, this type of scar time-crystallinity does not require coupling disorder and can occur in the absence of a Zeeman energy mismatch between central and satellite spins.

Our work also introduces a new type of TC called ``subspace scar time crystal'' in which a time-evolving subspace, rather than an individual quantum state, breaks the discrete time-translation symmetry of the Hamiltonian. In a recent theoretical work~\cite{Wampler_PRB_SubspaceTC2023}, the breaking of discrete time-translation symmetry by Floquet-Krylov (cellular automata) subspaces has been suggested, but not explicitly demonstrated in an example. Thus, our work demonstrates, to the best of our knowledge, the first stable subspace scar time-crystalline behavior without any disorder. 



The cental-spin systems considered in this work naturally arise in a variety of physical platforms, including color centers in solids, semiconductor quantum dots, and rare-earth ions, which are currently being explored for multiple applications within quantum information science and engineering such as quantum sensing, quantum networks, and quantum computing and simulation. It is interesting to consider whether time-crystallinity could be used to modify the functionality or enhance the performance of central-spin systems for such applications.

\section*{Acknowledgement}
This work was supported by the National Science Foundation (Grant No. 1847078) and by the Commonwealth Cyber Initiative (CCI), an investment in the advancement of cyber R\&D, innovation, and workforce development.

\appendix

\appendix
\section{Floquet-Krylov subspaces} \label{sec:Appen_Floq_Krylov}
In this appendix, we provide further details about the Krylov subspaces of the time-independent Hamiltonian $H_0$ and the Floquet-Krylov subspaces corresponding to the driven stroboscopic dynamics of the system (as discussed in Sec.~\ref{sec:Hilbert_Frag}).

\subsection{Krylov subspaces corresponding to unitary operator and Hamiltonian}
Claim 1: For a given state $\ket{\psi}$, the subspace spanned by a time evolution operator $U_{H_0}(t)=e^{-iH_0t}$ is equal to  the Krylov subspace of the Hamiltonian $H_0$, i.e.,
\begin{equation}
    \text{Span of } \big\{ U_{H_0}(t) \ket{\psi}\big\}_{t\in \mathbb{R}} =\mathcal{K}(H_0, \ket{\psi}),
\end{equation}
where, $\mathcal{K}(H_0, \ket{\psi})$ is defined in Eq.~(\ref{eq:Krylov_Sub_H0}).\\
 Proof: We can write the unitary operator as $e^{-iH_0t}=\sum_{n=0}^{\infty} \alpha^n H_0^n$, where $\alpha^n=(-it)^n/n!$ is either purely real or purely imaginary for an arbitrary time and integer $n$. Applying this to an arbitrary state $\ket{\psi}$, we have
 \begin{align}\label{eq:Krylov_H}
     U_{H_0}(t) \ket{\psi} &= \sum_{n=0}^{\infty} \alpha^n H_0^n \ket{\psi}  \nonumber\\
                    & = \alpha^0\ket{\psi} + \alpha^1 H_0 \ket{\psi} + \alpha^2 H_0^2 \ket{\psi}  +... \nonumber\\
                    & \hspace{10mm} +\alpha^n H_0^n \ket{\psi} +...
 \end{align}
Here, we see that  $\big\{ U_{H_0}(t) \ket{\psi}\big\}_{t\in \mathbb{R}}$ generates a set $\big\{\ket{\psi}, H_0 \ket{\psi}, H_0^2 \ket{\psi}, ..., H_0^n \ket{\psi},... \big\}$. But,  since $\alpha^{n} $ belongs to a subset of the complex numbers, it is not the same as the full Krylov subspace defined with arbitrary complex coefficients. However, since the evolution operator generates a continuous set of Krylov subspaces parameterized by $t$, if we consider arbitrary linear combinations across this set with complex coefficients, we have
 \begin{equation}
    \text{Span of } \big\{ U_{H_0}(t) \ket{\psi}\big\}_{t\in \mathbb{R}} =\mathcal{K}(H_0, \ket{\psi}).
\end{equation}

Let's consider an example to illustrate the above point more transparently. We take $H_0=\sigma_x$ and choose $\ket{\Psi}= \ket{\uparrow}
 $ so that $U_{H_0}(t)\ket{\Psi}= \cos t \ket{\uparrow} -i\sin t\ket{\downarrow}
 $. Here, even for arbitrary time $t$, we cannot generate any state $\ket{\chi}=c_1\ket{\uparrow}+c_2\ket{\downarrow}$, where $c_1,c_2$ are arbitrary complex numbers (with $|c_1|^2+|c_2|^2=1$). However, if we take linear combinations of states $U_{H_0}(t)\ket{\Psi}$ with different values of $t$, then it is possible to obtain any state $\ket{\chi}$. For example, we could form linear combinations of $U_{H_0}(0)\ket{\Psi}$ and $U_{H_0}(\pi/2)\ket{\Psi}$.

Thus, we see that a dynamical subspace spanned by an evolution operator is equal to the Krylov subspace associated with the Hamiltonian that generates it only when we consider arbitrary time. If the time is chosen to be some finite set (or countably infinite) then the dynamical subspace is a subset of the Krylov subspace in general. This will be more transparent when we look into the Floquet Hamiltonian and stroboscopic dynamics later in this section. 

 Claim 2: A subspace spanned by a stroboscopic evolution operator $\{ U(nT)\}_{n\in\mathbb{N}},\ket{\psi})$ is a subset of the Krylov subspace for the Floquet Hamiltonian, i.e.,
\begin{equation}
    \mathcal{K}_F(U_F, \ket{\psi})  \subseteq \mathcal{K}(H_F, \ket{\psi}),
\end{equation}
where $\mathcal{K}_F(U_F, \ket{\psi})=\text{Span of } \big\{ \{U(nT)\ket{\psi}\}_{n\in\mathbb{N}}\big\}$ and $\mathcal{K}(H_F, \ket{\psi})$ is defined as in Eq.~(\ref{eq:Krylov_Sub_H0}).  \\
 Proof: 
 \begin{align}
\mathcal{K}\big( H_F, \ket{\psi} \big) &=  \text{Span of } \big\{ \ket{\psi}, H_F\ket{\psi},...,H_F^n\ket{\psi},.. \big\}\nonumber\\
&=  \text{Span of } \big\{ e^{-iH_F t}\ket{\psi} \big\}_{t\in\mathbb{R}},\\
\mathcal{K}_F(U_F, \ket{\psi}) &=  \big\{ e^{-iH_F nT}\ket{\psi} \big\}_{n\in\mathbb{Z}}.
 \end{align}
 Here, time forms a countable set (integer multiples of the period) in $\mathcal{K}_F(U_F, \ket{\psi}$, but (real) time is uncountable for $\mathcal{K}\big( H_F, \ket{\psi} \big)$. So, $ \mathcal{K}_F(U_F, \ket{\psi})  \subseteq \mathcal{K}(H_F, \ket{\psi})$.

We can also illustrate this claim using the following example of the time-dependent Hamiltonian $ H(t) = \sum_{n\in \mathbb{Z}} \delta(t-nT) \pi(s_1^x + s_2^x)$, where, $s_i^x= \sigma_i^x/2$. We obtain
\begin{align}
    U_F &= e^{-i \pi(s_1^x + s_2^x)}  = -4 s_1^x s_2^x,\\
    H_F &= \frac{\pi}{T}(s_1^x + s_2^x).
\end{align}
When we consider the initial state $\ket{\uparrow \uparrow}$ then 
\begin{align}
    \mathcal{K} \big( H_F, \ket{\uparrow \uparrow} \big) &= \text{Span of }\big\{\ket{\uparrow \uparrow}, \ket{\downarrow \uparrow},\ket{\uparrow \downarrow},\ket{\downarrow \downarrow} \big\}, \\
    \mathcal{K}_F \big( U_F, \ket{\uparrow \uparrow} \big) &=\text{Span of }\big\{\ket{\uparrow \uparrow}, \ket{\downarrow \downarrow}\big\}. 
\end{align}
This example demonstrates that, in general, $\mathcal{K}_F(U_F, \ket{\psi})  \neq \mathcal{K}(H_F, \ket{\psi})$, but  
$ \mathcal{K}_F(U_F, \ket{\psi})  \subseteq \mathcal{K}(H_F, \ket{\psi})$.

\subsection{Floquet-Krylov subspaces of a driven homogeneous central-spin Hamiltonian}
In this appendix, by analyzing the the repeated action of $U_F$, we obtain the Floquet-Krylov subspace for initial state $\ket{m_j\sigma}$. As discussed in Sec.~\ref{sec:Hilbert_Frag} of the main text, the Krylov subspace generated by the time-independent Hamiltonian $H_0$ is 
\begin{align}
    \mathcal{K}\big(H_0, \ket{J\uparrow}\big) &= \text{Span of }\big\{ \ket{J\uparrow}\big\}, \\
    \mathcal{K}\big(H_0, \ket{m_j\downarrow}\big) &= \text{Span of }\big\{ \ket{(m_j-1)\uparrow}, \ket{m_j\downarrow} \big\}, \\
    \mathcal{K}\big(H_0, \ket{m_j\uparrow}\big) &= \text{Span of }\big\{ \ket{m_j\uparrow}, \ket{(m_j+1)\downarrow} \big\}.
\end{align}

Next, we calculate the action of $U_\pi$, where $U_F= U_{\pi}U_{H_0}$, on the basis of $\mathcal{K}(H_0, \ket{m_j\sigma}$. We decompose $U_\pi$ in terms of individual spin-$\frac{1}{2}$ operators:
\begin{align}
    U_\pi =e^{-i\pi (\sum_{j} S_j^x + S_0^x)} &=  \prod_{j=1}^{N} e^{-i\pi S_j^x} e^{-i\pi S_0^x}\nonumber \\
    &=   (-i)^{N+1}\prod_{j=1}^{N}\sigma_j^x \sigma_0^x.
\end{align}
The satellite part $\ket{m_j}$ can be written as a superposition of local $z$-basis product states as $  \ket{m_j} =\sum_\alpha c_\alpha \ket{\sigma_1\sigma_2...\sigma_j...\sigma_N}_\alpha$ where $\ket{\sigma_j}\in \{ \ket{\uparrow},  \ket{\downarrow}\}$. Any state $\ket{\sigma_1\sigma_2......\sigma_j....\sigma_N}$ in the superposition has same the number of up spins ($N_{\uparrow}$) and down spins ($N_{\downarrow}$) such that $m_j= (N_{\uparrow} -N_{\downarrow})/2 $.  So,
\begin{align}
    U_\pi \ket{\sigma_1\sigma_2...\sigma_j...\sigma_N} \ket{\sigma_0} = \ket{\bar\sigma_1\bar\sigma_2...\bar\sigma_j...\bar\sigma_N} \ket{\bar\sigma_0},
\end{align}
 where $\ket{\bar{\sigma}}=\sigma^x\ket{\sigma}$. Thus, $ U_\pi \ket{m_j\sigma}= \ket{-m_j\bar{\sigma}}$.

 Now we take $\ket{\sigma_0}= \ket{\downarrow}$ (without loss of generality) and apply $U_F$ on the state $\ket{m_j \downarrow}$:
 \begin{align}
  U_{H_0} \ket{m_j \downarrow} \in \text {Span of }\{ \ket{m_j \downarrow},\ket{(m_j-1)\uparrow} \}, \nonumber \\
U_F \ket{m_j \downarrow} \in \text {Span of } \{ \ket{-m_j \uparrow}, \ket{-(m_j-1) \downarrow}\}.
 \end{align}
 Now if we apply $U_F$ again, we have 
  \begin{align}
    U_{H_0} U_F \ket{m_j \downarrow} & \in \text {Span of } \{\ket{-m_j \uparrow}, \ket{-m_j+1 \downarrow}\}, \nonumber \\
      U_F^2 \ket{m_j \downarrow} &\in \text {Span of } \{ \ket{m_j \downarrow}, \ket{(m_j-1) \uparrow}\}.
 \end{align}
 Here, it is easy to see that even if we had started with $\ket{(m_j-1) \uparrow}$, we would find the same subspaces. 
 Moreover, as we calculate the repeated action of $U_F$ on the initial state $\ket{m_j \downarrow}$, all the odd multiples of $U_F$ produce a 2D subspace spanned by $\{\ket{-m_j \uparrow}, \ket{-m_j+1 \downarrow}\}$, while even multiples produce $\{ \ket{m_j \downarrow}, \ket{(m_j-1) \uparrow}\}$.  Thus, we obtain a 4D Floquet-Krylov subspace as given in Eq.~(\ref{eq:Floq_Krylov2}) of the main text. Similarly, we can choose the initial state $\ket{m_j \uparrow}$ and obtain the  Floquet-Krylov subspace as given in Eq.~(\ref{eq:Floq_Krylov1}).

\subsection{Growth of fragmentated space over time}
\begin{figure}[t] 
   \includegraphics[width=3.4in]{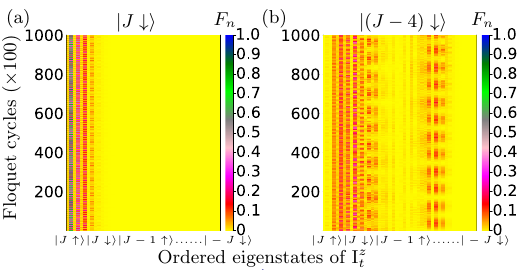}
  \caption{The overlap of the time-evolved state, starting from initial state (a) $\ket{J\downarrow}$ or (b) $\ket{(J-4)\downarrow}$, with each eigenstate of the total satellite spin operator $I^z_{t}$ for a periodically driven Heisenberg central-spin model. Results for up to $10^5$ Floquet cycles in steps of $100$ cycles are shown. The parameters are $N=2J=21$, $A=1.3$ MHz, $B_z=100$ MHz, $\omega=1$ MHz, $\theta=0.1\pi$. The color bar shows the overlap $F_n(\ket{m_j\sigma},\ket{\psi})$ where $\ket{\psi}=\ket{J\downarrow},\ket{(J-4)\downarrow}$. }
  \label{fig:Krylov_Growth_Jdown}
\end{figure}

 Here, we discuss the growth of the Floquet-Krylov subspaces (fragmented spaces) with respect to time when the pulse error is small. We consider the driven Heisenberg central-spin model and numerically calculate the overlaps $F_n$ (as  defined in Eq.~(\ref{eq:overlap})) of the evolved state with respect to each basis state. In Fig.~\ref{fig:Krylov_Growth_Jdown}, we show the 
overlap with each eigenstate of $\I_t^z$ for up to $10^5$ Floquet cycles (results after every 100 cycles are shown).
Fig.~\ref{fig:Krylov_Growth_Jdown}(a) shows that the return probability for the initial state $\ket{J\downarrow}$ is high (around 0.7), and that only overlaps with neighboring states become significant. This suggests that, even on a long time scale, the system is constrained to highly polarized states and shows non-ergodic behavior. In contrast, for the initial state $\ket{(J-4)\downarrow}$ (Fig.~\ref{fig:Krylov_Growth_Jdown}(b)), the overlaps with a large number of eigenstates of $\I_t^z$ become significant, including states with negative eigenvalues. Therefore, the system shows ergodic behavior for the observable $\I^{z}_t$ when it is initialized in the state $\ket{(J-4)\downarrow}$.

\section{Exact Stroboscopic dynamics under Ising and XX central-spin interactions}\label{sec:Appen_Exact_Ising_XX}

In this appendix, we derive analytical expressions for various dynamical quantities (without and with pulse errors) for periodically driven central-spin models with Ising and XX interactions. In both cases, the initial state is $\ket{m\sigma}$, and we include a Zeeman field for the Ising case but not the XX case. 

We can write the Floquet operator with pulse error $\theta$ as 
\begin{align}
    U(T,\theta) &= U_F(\theta) = U_{\theta}U_{\pi}U_{H_0}, \nonumber  \\
    U_{\theta} &\approx \mathbb{1}-\frac{i\theta}{2} P+\frac{-\theta^2}{4} P^2 + O(\theta^3),
\end{align}
where  $ P=(\I_t^{+}+\I_t^{-}+\sigma_0^x)$. Up to $O(\theta)$, we can write the  time evolution after $n$ periods as 
\begin{align}
    U(nT,\theta) &= U_F^n -\frac{i\theta}{2}  \sum_{j=0}^{n-1} U_F^j P U_F^{n-j} +O(\theta^2),
\end{align}
For even cycles ($2p, p \in \mathbb{N} $), the Floquet operator becomes
\begin{equation} \label{eq:UF_theta1}
    U(2pT,\theta) \approx U_F^{2p} -\frac{i\theta}{2} \Big(\sum_{r=1}^{2p-1} U_F^r P U_F^{2p-r} +P U^{2p}_F \Big) +O(\theta^2).
\end{equation}
Before continuing, it helps to write out the action of $\I_t^{\pm}$ on an eigenstate of $\I_t^z$ in the largest symmetry sector of $\I_t^2$  ($j=J=N/2$) and denote $\ket{m_J\sigma}= \ket{m\sigma}$. We have
\begin{align} \label{eq:I_t_plus_minus}
    \I^{\pm}_t\ket{m\sigma}= \alpha_m^{\pm}\ket{m\pm 1\sigma},
\end{align}
where $\alpha_m^{\pm}=\sqrt{N(N+2)/4 - m(m\pm 1)}$, and we denote $\alpha_m^0=\alpha_m^+$. We write some relations of $\alpha_m^{\pm}$ as
\begin{align}
    \alpha_{-m}^{+} &= \alpha_{m-1}^{+}=  \alpha_{m-1}^{0}, \nonumber\\
     \alpha^{-}_{\pm m} &=  \alpha^{+}_{\mp m}.
\end{align}
\subsection{Ising type Hamiltonian} \label{sec:Appen_Ising_Exact_dynamics}
Now we consider $H_0=H_z= A_z \I_t^zS_0^z +  B_z S_0^z$ and  $U_{H_z}=e^{-iH_z T}$. 
\begin{align}
    U_F\ket{m\sigma} &= e^{-i\Phi_m^{\sigma}}\ket{-m\bar{\sigma}}, \nonumber  \\
 U^2_F\ket{m\sigma} &= e^{-i\Phi_m^{\sigma}}e^{-i\Phi_{-m}^{\bar{\sigma}}}\ket{m\sigma}= e^{-i(A_zmT\sigma)}\ket{m\sigma}, 
 \end{align}
 where $\Phi_m^{\sigma}= (A_zm +B_z)T\sigma/2$ and $\sigma =1(-1)$ for state $\ket{\uparrow} (\ket{\downarrow})$. In the last line of the above equation, we have used  $\Phi_m^{\sigma} +\Phi_{-m}^{\bar{\sigma}}=A_zmT\sigma$. So, after even and odd Floquet cycles, the initial state $\ket{m\sigma}$ evolves as 
\begin{align}
    U^{2p}_F\ket{m\sigma} &=  e^{-ip(A_zmT\sigma)}\ket{m\sigma}, \\
     U^{2p+1}_F\ket{m\sigma} &=  e^{-i(pA_zmT\sigma+\Phi_m^{\sigma})}\ket{-m\bar{\sigma}}. 
\end{align}

Here, for both odd and even cycles, the phase factors do not depend on the number of satellite spins, and for even cycles, they also do not depend on the central-spin Zeeman field.
Now we look into the effect of small pulse errors on the evolution of $\ket{m\sigma}$ after $2p$ cycles using  Eq.~(\ref{eq:UF_theta1}). We split the summation in  Eq.~(\ref{eq:UF_theta1}) into two parts: (i) $r$ and $2p-r$ are even, (ii) $r$ and $2p-r$ are odd. For convenience, we consider $\ket{\sigma}= \ket{\uparrow}$. First we have  
 \begin{align} \label{eq:UF_Ising_theta1}
  PU^{2p}_F\ket{m\uparrow} &= e^{-ipA_zmT}\big[\alpha_m^+\ket{m+1\uparrow} + \nonumber\\
 &\hspace{20mm}\alpha_m^- \ket{m-1\uparrow} + \ket{m\downarrow}\big]. 
 \end{align}
 Now we consider the case where $r$ and $2p-r$ are even in the summation of  Eq.~(\ref{eq:UF_theta1}).
 \begin{align}\label{eq:UF_Ising_theta2}
 \sum_{\substack{r=2q,\\q\in \mathbb{N}}}^{2p-2}U_F^{r}PU_F^{2p-r}\ket{m\uparrow}&= \sum_{\substack{r=2q,\\q\in \mathbb{N}}}^{2p-2} U_F^r\big[e^{-i(2p-r)A_zmT/2}\times \nonumber\\
 &\hspace{-10mm}\{\alpha_{m}^{+}\ket{m+1\uparrow} +\alpha_{m}^{-}\ket{m-1\uparrow}+\ket{m\downarrow}\}\big] \nonumber \\
      &\hspace{-20mm}=e^{-ipA_zmT}\sum_{r=2q,q\in \mathbb{N}}^{2p-2}\Big[\alpha_{m}^0e^{-irA_zT/2}\ket{m+1\uparrow} +\nonumber\\
      &\hspace{-20mm }\alpha_{m-1}^0e^{irA_zT/2}\ket{m-1\uparrow}+ e^{irA_zmT}\ket{m\downarrow}\Big].
\end{align}
Now we consider the case where $r$ and $2p-r$ are odd in the summation of  Eq.~(\ref{eq:UF_theta1}).
 \begin{align}\label{eq:UF_Ising_theta3}
 &\hspace{-3mm}\sum_{\substack{r=2q-1,\\ q\in \mathbb{N}}}^{2p-1}U_F^{r}PU_F^{2p-r}\ket{m\uparrow}= \sum_{\substack{r=2q-1,\\ q\in \mathbb{N}}}^{2p-1} U_F^r\big[e^{-i(\Phi_{m}^{\uparrow}+\frac{2p-r-1}{2}A_zmT)} \nonumber\\
 &\hspace{8mm}\times\{\alpha_{-m}^{+}\ket{-m+1\downarrow} +\alpha_{-m}^{-}\ket{-m-1\downarrow}+\ket{-m\uparrow}\}\big] \nonumber \\
      &\hspace{4mm}=e^{-ipA_zmT}\sum_{r=2q-1,q\in \mathbb{N}}^{2p-1}\Big[\alpha_{m}^0e^{-irA_zT/2}\ket{m+1\uparrow} +\nonumber\\
      &\hspace{4mm }\alpha_{m-1}^0e^{irA_zT/2}\ket{m-1\uparrow}+ e^{i(rA_zmT-B_zT)}\ket{m\downarrow}\Big].
\end{align}
Thus, using Eqs.~ (\ref{eq:UF_Ising_theta1},\ref{eq:UF_Ising_theta2},\ref{eq:UF_Ising_theta3}),  we  have up to  ${\cal O}(\theta)$:,
\begin{align}\label{eq:Ising_pertubation}
U(2pT,\theta)\ket{m\uparrow} &= \ket{m\uparrow} -i\frac{\theta}{2} e^{-ipA_zmT} \big[\{1 + \nonumber\\
&\hspace{-15mm}\sum_{\substack{r=2q,\\ q\in \mathbb{N}}}^{2p-2}e^{irA_zmT}+\sum_{\substack{r=2q-1,\\q\in \mathbb{N}}}^{2p-1}e^{i(rA_zmT-B_zT)}\}\ket{m\downarrow}+ \nonumber\\
&\alpha_m^0\{1+\sum_{r=1}^{2p-1}e^{-irA_zT/2}\}\ket{m+1\uparrow}  + \nonumber\\
&\hspace{-5mm}\alpha_{m-1}^0\{1+\sum_{r=1}^{2p-1}e^{irA_zT/2}\}\ket{m-1\uparrow}\big].
\end{align}
The system becomes more robust against $\theta$ when it returns to the same satellite spin state. Thus, the coefficients of states $\ket{(m\pm 1)\sigma}$ should be zero, i.e., 
\begin{align}
   (1+\sum_{r=1}^{2p-1}e^{\pm irA_zT/2})&=(1+\sum_{r=1}^{2p-1}e^{\pm ir(A_z/\omega)\pi})=0\nonumber\\  
   &\Rightarrow A_z/\omega= (2p+1),
\end{align}
where $p\in \mathbb{Z}$. Thus from here, we see that when $A_z $ is an odd multiple of the drive frequency, we get a more robust time-crystalline behavior.  In the main text, we considered $\ket{\psi(0)}=\ket{J\uparrow}$ in Sec.~\ref{sec:phase_diagm}. We also chose $J=N/2=10.5$ and integer values of $B_z/\omega$. In that case and for $A_z/\omega$ an odd integer, the coefficient of $\ket{m\downarrow}=\ket{J\downarrow}$ (as in Eq.~(\ref{eq:Ising_pertubation})) is also zero. So, after $2p$ cycles, we obtain
\begin{align}
   &\ket{\psi(2pT,\theta)}=\ket{J\uparrow} +O(\theta^2), \\
    &\sandwich{\psi(2pT,\theta)}{\I_t^z}{\psi(2pT,\theta)}= J + O(\theta^2).
\end{align}
This analytical result agrees with the numerical results shown in Fig.~\ref{fig:fig6_ScarTC_phase_theta_A}(a).

\subsection{XX type Hamiltonian with zero Zeeman energy}
Here, we consider an XX Hamiltonian with zero Zeeman energy:
$H_0= H_{xx}= A(\I_t^{+}S_0^{-}+ \I_t^{-}S_0^{+})$. We have
\begin{align}
    U_{H_{xx}}\ket{m\uparrow} &= (1 -iTH_{xx} +\frac{(-iT)^2}{2!}H^2_{xx}+\frac{(-iT)^3}{3!}H^3_{xx}  \nonumber\\
    &\hspace{40mm} + ....)\ket{m\uparrow} \nonumber\\
    & \hspace{-13mm}=\cos(AT\alpha_m^0)\ket{m\uparrow} -i\sin(AT\alpha_{m}^0)\ket{(m+1)\downarrow}, \nonumber\\
    U_{H_{xx}}\ket{m\downarrow} &= \cos(AT\alpha_{m-1}^0)\ket{m\downarrow} -i\sin(AT\alpha_{m-1}^0) \nonumber\\
    &\hspace{40mm}\ket{(m-1)\uparrow},  \nonumber
\end{align}

Here  $\alpha_{m}^0=\alpha_{m}^+$ is defined in Eq.~(\ref{eq:I_t_plus_minus}). 
We denote $C_m=\cos(AT\alpha_m^0)$, $S_m=\sin(AT\alpha_m^0)$. Now we have
    \begin{align}
     U_F\ket{m\downarrow} & = C_{m-1}\ket{-m\uparrow} -i S_{m-1} \ket{(-m+1)\downarrow}, \nonumber  \\
    U_F\ket{m\uparrow} & = C_m\ket{-m\downarrow} -i S_m \ket{(-m-1)\uparrow}, \nonumber  \\
    U^2_F\ket{m\uparrow} &=U_{\pi}\big[ C_m[C_{-m-1}\ket{-m\downarrow} -iS_{-m-1}\ket{-m-1\uparrow}]  \nonumber\\
    &\hspace{-5mm}-i S_m[ C_{-m-1}\ket{(-m-1)\uparrow}-i S_{-m-1}\ket{-m\downarrow}]\big]
    \nonumber\\
    &=(C_mC_{-m-1} -S_m S_{-m-1})\ket{m\uparrow}  \nonumber\\
    & -i(S_m C_{-m-1} + C_m S_{-m-1})\ket{(m+1)\downarrow} \nonumber\\
    &\hspace{-7mm}=(C_m^2 -S_m^2)\ket{m\uparrow} -i2S_m C_{m}\ket{(m+1)\downarrow}.
\end{align}
Similarly, we can find
 \begin{align}
    U^2_F\ket{m+1\downarrow}=&(C_m^2 -S_m^2)\ket{m+1\downarrow} -i2S_m C_{m}\ket{m\uparrow}.
\end{align}
We denote $\beta^m_2= (C_m^2 -S_m^2)$, $\gamma^m_2= -i2C_mS_m$. After four Floquet cycles, we have
\begin{align}
    U^4_F\ket{m\uparrow}&= ((\beta_2^m)^2+(\gamma_2^m)^2)\ket{m\uparrow} + 2\beta^m_2\gamma^m_2\ket{(m+1)\downarrow}. \nonumber
\end{align}
We denote $\beta_4^m= ((\beta_2^m)^2+(\gamma_2^m)^2)$, $\gamma_4= 2\beta_2^m\gamma_2^m$. After six Floquet cycles, we obtain
   \begin{align} 
     U^6_F\ket{m\uparrow}&= (\beta_4^m\beta_2^m+\gamma_4^m\gamma_2^m)\ket{m\uparrow} +\nonumber\\
     &\hspace{10mm} (\beta_4^m\gamma_2^m+ \gamma_4^m\beta_2^m)\ket{m+1\downarrow}. \nonumber
\end{align}
We denote $\beta_6^m= (\beta_4^m\beta_2^m+\gamma_4^m\gamma_2^m)$, $\gamma_6^m=(\beta_4^m\gamma_2^m+ \gamma_4^m\beta_2^m)$. In general, for even ($2p, p\in \mathbb{N}$) Floquet cycles, we obtain
\begin{align}
     U^{2p}_F\ket{m\uparrow}&= \beta_{2p}^m\ket{m\uparrow} +  \gamma_{2p}^m\ket{(m+1)\downarrow}, \\
     &\hspace{-22mm}\text{where, for } p\geq 2,   \nonumber\\
     \beta_{2p}^m &=(\beta_{2p-2}^m\beta_2^m+\gamma_{2p-2}^m\gamma_2^m),\\
     \gamma_{2p}^m &= (\beta_{2p-2}^m\gamma_2^m+\gamma_{2p-2}^m\beta_2^m).
\end{align}
Similarly, for odd cycles, we have
\begin{align}
     U^{2p+1}_F\ket{m\uparrow}&= U_F(\beta_{2p}\ket{m\uparrow} +  \gamma_{2p}\ket{(m+1)\downarrow}) \nonumber \\
     &\hspace{-5mm}=\beta_{2p+1}^m\ket{-m\downarrow} +  \gamma_{2p+1}^m\ket{-m-1\uparrow}, \\
     &\hspace{-22mm}\text{where} \nonumber\\
     \beta_{2p+1}^m &=(\beta_{2p}^mC_m-i\gamma_{2p}^mS_m),\\
     \gamma_{2p+1}^m &= (\gamma_{2p}^mC_m-i\beta_{2p}^mS_m).
\end{align}

Here, the exact evolution of the state $\ket{m\uparrow}$, after even cycles, is constrained to be a linear combination of $\ket{m\uparrow},\ket{m+1\downarrow}$  and, after odd cycles, to lie in a subspace  spanned by $\ket{-m\downarrow},\ket{-m-1\uparrow}$.  The coefficients depend on the number of satellite spins, the total satellite spin quantum number $m$, and the effective coupling terms $AT$.  Unlike the Ising case, here these parameters affect the magnitude of the coefficients.

Now we find the evolution after two Floquet cycles, for small pulse error, up to ${\cal O}(\theta)$:
\begin{align}
   U_FPU_F\ket{m\uparrow} &=U_F\big[C_m P\ket{-m\downarrow} -i S_m P\ket{(-m-1)\uparrow}\big] \nonumber  \\
   &\hspace{-20mm}=U_F\big[C_m \big(\alpha_{-m}^+\ket{-m+1\downarrow}+\alpha_{-m}^-\ket{-m-1\downarrow}+ \ket{-m\uparrow}\big)  \nonumber  \\
   &\hspace{-20mm}-i S_m \big(\alpha_{-m-1}^+\ket{-m\uparrow}+\alpha_{-m-1}^-\ket{-m-2\uparrow}+ \ket{-m-1\downarrow}\big)\big]  \nonumber\\
 &\hspace{-20mm}=\big[C_m \big(\alpha_{-m}^+\{C_{-m}\ket{m-1\uparrow}-iS_{-m}\ket{m\downarrow}\} \nonumber\\
    &\hspace{-15mm}+\alpha_{-m}^-\{C_{-m-2}\ket{m+1\uparrow} 
    -iS_{-m-2}\ket{m+2\downarrow}\}  \nonumber\\
    &\hspace{-15mm}+\{C_{-m}\ket{m\downarrow} -iS_{-m}\ket{m-1\uparrow}\}\big)\big] + \nonumber\\
 &\hspace{-15mm}\big[-iS_m \big(\alpha_{-m-1}^+\{C_{-m}\ket{m\downarrow}-iS_{-m}\ket{m-1\uparrow}\} \nonumber\\
    &\hspace{-10mm}+\alpha_{-m-1}^-\{C_{-m-2}\ket{m+2\uparrow} 
    -iS_{-m-2}\ket{m+1\uparrow}\}  \nonumber\\
    &\hspace{-15mm}+\{C_{-m-2}\ket{m+1\uparrow} -iS_{-m-2}\ket{m+2\downarrow}\}\big)\big] \nonumber\\
&\hspace{-20mm}= \big[C_m(C_{m-1}\alpha_{m-1}^0-iS_{m-1}) -S_mS_{m-1}\alpha_{m}^0\big]\ket{m-1\uparrow} + \nonumber \\
&\hspace{-20mm}\big[C_m(-iS_{m-1}\alpha_{m-1}^0 + C_{m-1}) -iS_mC_{m-1}\alpha_{m}^0\big]\ket{m\downarrow}+  \nonumber \\
&\hspace{-20mm}\big[C_mC_{m-1}\alpha_{m}^0-S_m(S_{m+1}\alpha_{m}^0 +iC_{m+1})\big]\ket{m+1\uparrow} +   \nonumber \\
&\hspace{-20mm}\big[-iC_mS_{m}\alpha_{m}^0-S_{m} (S_{m+1}+iC_{m+1}\alpha_{m+1}^0)\big]\ket{m+2\downarrow},
 \end{align}
 \begin{align}
 PU^2_F\ket{m\uparrow} &=(C_m^2 -S_m^2)[\alpha_m^0\ket{m+1\uparrow} + \alpha_{m-1}^0\ket{m-1\uparrow}+ \nonumber\\  
 &\hspace{-15mm}\ket{m\downarrow}] -i2S_m C_m[\alpha_{m+1}^0\ket{m+2\downarrow} +\alpha_{m}^0\ket{m\downarrow} +\ket{m+1\uparrow}].
\end{align}
So, we have
\begin{align}
    U(2T,\theta)\ket{m\uparrow} &= G_m^{\uparrow}\ket{m\uparrow}+ G_{m+1}^{\downarrow}\ket{m+1\downarrow} -\frac{i\theta}{2} \big( G_m^{\downarrow}\ket{m\downarrow} \nonumber\\
    &\hspace{-20mm}+ G_{m-1}^{\uparrow}\ket{m-1\uparrow}+ G_{m+1}^{\uparrow}\ket{m+1\uparrow}  
+G_{m+2}^{\downarrow}\ket{m+2\downarrow} \big),
\end{align}
where we have defined
\begin{align}
    G_m^{\uparrow}&= (C_m^2 -S_m^2),\nonumber\\
    G_{m+1}^{\downarrow}&= -i2S_m C_{m},\nonumber\\
    G_m^{\downarrow} &= 1-C_m(-iS_{m-1}\alpha_{m-1}^0 + C_{m-1})  - \nonumber\\ &\hspace{10mm}iS_mC_{m-1}\alpha_{m}^0+\alpha_m^0,\nonumber\\
     G_{m+1}^{\uparrow} &=C_mC_{m-1}\alpha_{m}^0-S_m(S_{m+1}\alpha_{m}^0 +iC_{m+1}) + 2C_m^2,\nonumber\\
    G_{m-1}^{\uparrow} &= C_m(C_{m-1}\alpha_{m-1}^0-iS_{m-1}) -S_mS_{m-1}\alpha_{m}^0 +\alpha_{m-1}^0,\nonumber\\
    G_{m+2}^{\downarrow} &= -iC_mS_{m}\alpha_{m}^0-S_{m} (S_{m+1}+iC_{m+1}\alpha_{m+1}^0)\nonumber\\
    & \hspace{10mm}-i2S_m C_m\alpha_{m+1}^0.
\end{align}

Here, the presence of the pulse error $\theta$ causes the evolved state to overlap some states close to the initial state. Apart from $\theta$, the exact coefficients of the states also depend on the number of satellite spins, the total spin quantum number $m$, and the effective couplings.

\section{Floquet scar for anisotropic Heisenberg (XXZ) interactions} \label{sec:Appen_Floq_Scar_XXZ}
\begin{figure}[h]
   \centering
    \includegraphics[width=3.6in]{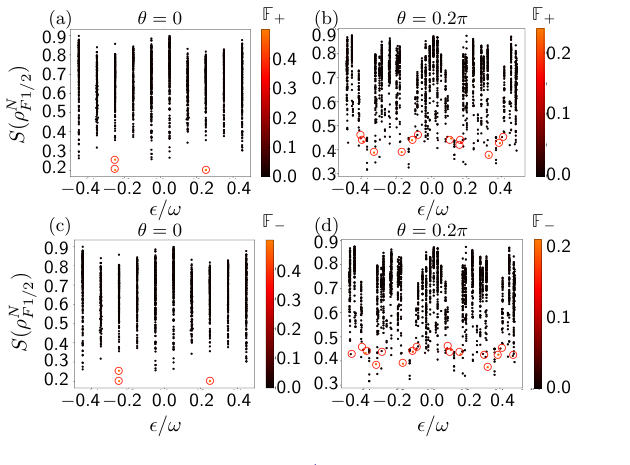}      
  \caption{
  Bipartite entanglement entropy of each Floquet eigenstate (labeled by its quasienergy $\epsilon/\omega$) in the satellite subsystem ($\rho^N_F$) and its overlap $\mathbb{F}_{\pm}$ with the fully polarized satellite states for a periodically driven central-spin model with XXZ interactions. The color bar shows the overlap $\mathbb{F}_{\pm}$ of the fully polarized satellite spin states $\ket{\pm J}$ with the satellite part of each Floquet eigenstate $\rho^N_F$. Here, $N=10=2J$, $A_{xy}$=1.3 MHz, $A_z$=0.4 MHz, $B_z$= 100 MHz, $\omega$= 1 MHz, and $B^n_z$= 0 MHz. The pulse error is (a,c) $\theta=0$ or (b,d) $\theta=0.2\pi$. Here, $\mathbb{F}_{\pm}\equiv\mathbb{F}(\ket{\pm J},\rho^N_F)$ is defined in Eq.~(\ref{eq:FloquetOverlaps}).}
   \label{fig:Floq_Scar_xxz}
\end{figure}
In this appendix, we examine the scar states (Floquet eigenstates with atypically low bipartite entanglement entropy) of the driven XXZ central-spin system and their overlap with fully polarized satellite states ($\ket{\pm J}$), similar to the study done for the Heisenberg interaction in Sec.~\ref{sec:Scar_Diagonosis_XXX} of the main text. 

In Fig~~\ref{fig:Floq_Scar_xxz}, we show the bipartite entanglement entropy of the satellite spin subsystem for each Floquet eigenstate. The $x$-axis shows the quasienergies in increasing order. The color bars show the overlaps ($\mathbb{F}_{\pm}$) of fully polarized satellite states with the Floquet eigenstates. The overlap $\mathbb{F}_{\pm}$ is defined in Eq.~(\ref{eq:FloquetOverlaps}). The circled dots show the nonzero overlaps of the satellite part of the Floquet eigenstates with fully polarized satellite states. 

Fig~\ref{fig:Floq_Scar_xxz} (a,c) shows the unperturbed case ($\theta=0$) where $\ket{\pm J}$ have high overlap with satellite part of the Floquet eigenstates with atypically low entanglement entropy. On the other hand, Fig~\ref{fig:Floq_Scar_xxz} (b,d) shows the overlap when the pulse error is $20\%$. We see that the states $\ket{\pm J}$ have larger overlap with Floquet eigenstates that have entanglement entropy in the lower range of the entanglement spectrum. Thus, for XXZ central-spin interactions, fully polarized satellite states have high overlaps with Floquet eigenstates with atypically low bipartite entanglement entropy.
\section{Time-crystallinity in the XX central-spin model with coupling disorder} \label{sec:Appen_XX_random_bond}

 In this appendix, we study the robustness of the scar DTC behavior of the XX central-spin system without a Zeeman field in the presence of random coupling disorder. The disorder breaks the   $\I_t^2$ symmetry of the Floquet operator, and therefore we cannot decompose the Hilbert space into different symmetry sectors. Therefore, we  consider the full Hilbert space and the local spin-$z$ basis for the calculation of magnetization. We take a fully polarized satellite state ($\ket{J\downarrow}=\ket{\uparrow_1\uparrow_2\uparrow_3\uparrow_4\uparrow_5} \ket{\downarrow_0)}$ with 5 satellite spins and one central spin as the initial state. In this case, the average satellite spin magnetization is the same as the magnetization at each individual site. We sample the couplings from a Gaussian distribution with mean value $A_{xy}$ and variance (disorder strength) $\delta A_{xy}$.  
\begin{figure}[h] 
   \includegraphics[width=3.37in]{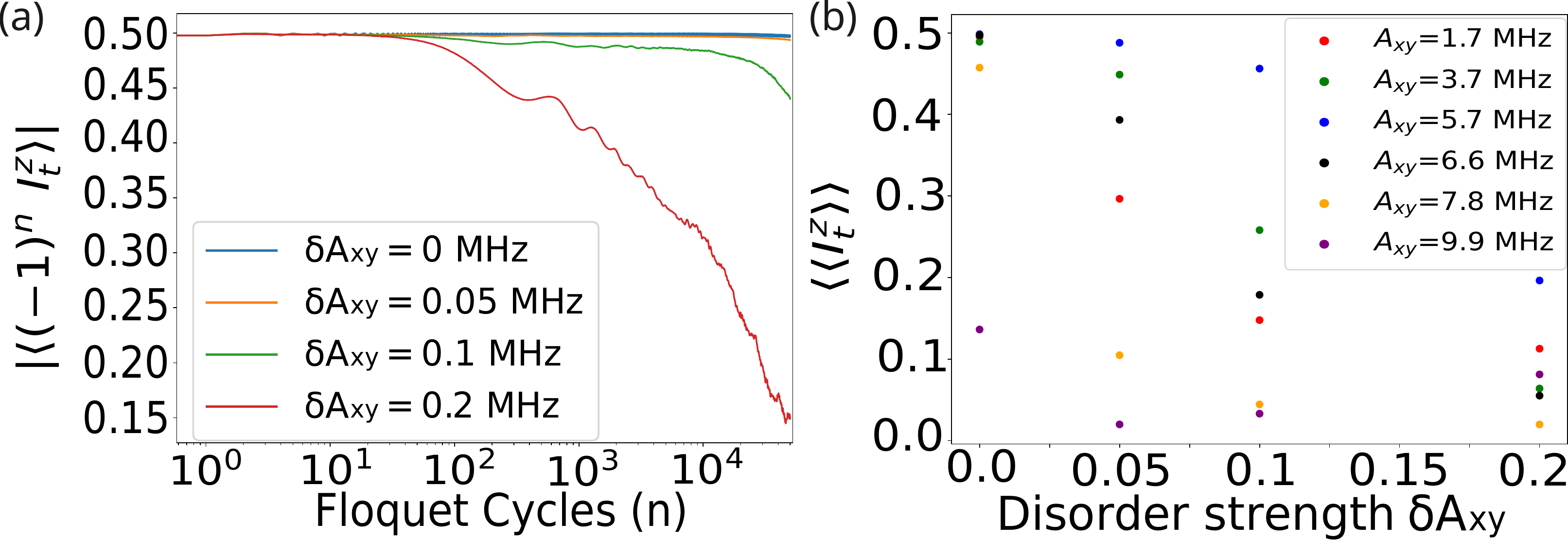}
  \caption{Effect of random couplings in the XX central-spin system without a Zeeman field.  The couplings are sampled from a Gaussian distribution with mean $A_{xy}$ and variance (disorder strength) $\delta A_{xy}$. (a) Staggered average satellite magnetization as a function of Floquet cycle for $A_{xy}=5.7$ MHz and for different values of $\delta A_{xy}$. (b) Time-averaged staggered satellite magnetization ($\bbraket{\I_t^z}$) as a function of the disorder strength $\delta A_{xy}$ for different values of the mean coupling $A_{xy}$. Here, $N=5$, the pulse error is $\theta=0.03\pi$, $\omega=1$ MHz, and the initial state is $\ket{J\downarrow}$.}
  \label{fig:j57scardisorder}
\end{figure}

In Fig.~\ref{fig:j57scardisorder} (a), we show the staggered average magnetization for 
the total satellite spin $\I_t^z$ over many Floquet cycles for different disorder coupling strengths. The average magnetization is about 0.5 in the absence of disorder (as expected from our study in the main text) or for a tiny amount of disorder across a large number of Floquet cycles. However, as the disorder strength is increased, the value of the average magnetization quickly goes to zero. Thus, disorder in the couplings adversely affects the robustness of time-crystalline behavior in the XX central-spin model.

In Fig.~\ref{fig:j57scardisorder} (b), we show the time average of the staggered magnetization ($\bbraket{\I_t^z}$) over $5\times 10^4$ Floquet cycles (as defined in Eq.(\ref{eq:time-averaged_magnetization})) as a function of the disorder strength ($\delta A_{xy}$). Different color dots for a given disorder strength represent the different values of the mean coupling $A_{xy}$. For the disorder-free  ($\delta A_{xy}=0$) case, we find that $\bbraket{\I_t^z}\approx 0.5$ for more values of $A_{xy}$. But as we increase the disorder strength, fewer values of $A_{xy}$ lead to $\bbraket{\I_t^z}\approx 0.5$. Moreover for $\delta A_{xy}=0.2$, none of the considered values of $A_{xy}$ exhibit $\bbraket{\I_t^z}\approx 0.5$. This further illustrates that an increase in the coupling disorder decreases the robustness of the time-crystalline behavior in the driven XX central-spin system.
\section{Robustness of DTC in XXX and XXZ with magnetic field  presence of coupling disorder}
In this appendix, we explore the time-crystalline behavior of a fully polarized nuclear spin in the presence of disordered couplings for both the Heisenberg (XXX) and XXZ models, with a magnetic field applied to the central spin. The introduction of disorder renders the system non-integrable. The results obtained in this section can be interpreted as addressing two possible questions: (a) whether the time-crystalline behavior observed in the integrable system persists when the system becomes non-integrable, and (b) whether the time-crystalline behavior found under homogeneous (uniform) coupling persists in the presence of random inhomogeneities.

\begin{figure} [h]
   \includegraphics[width=3.37in]{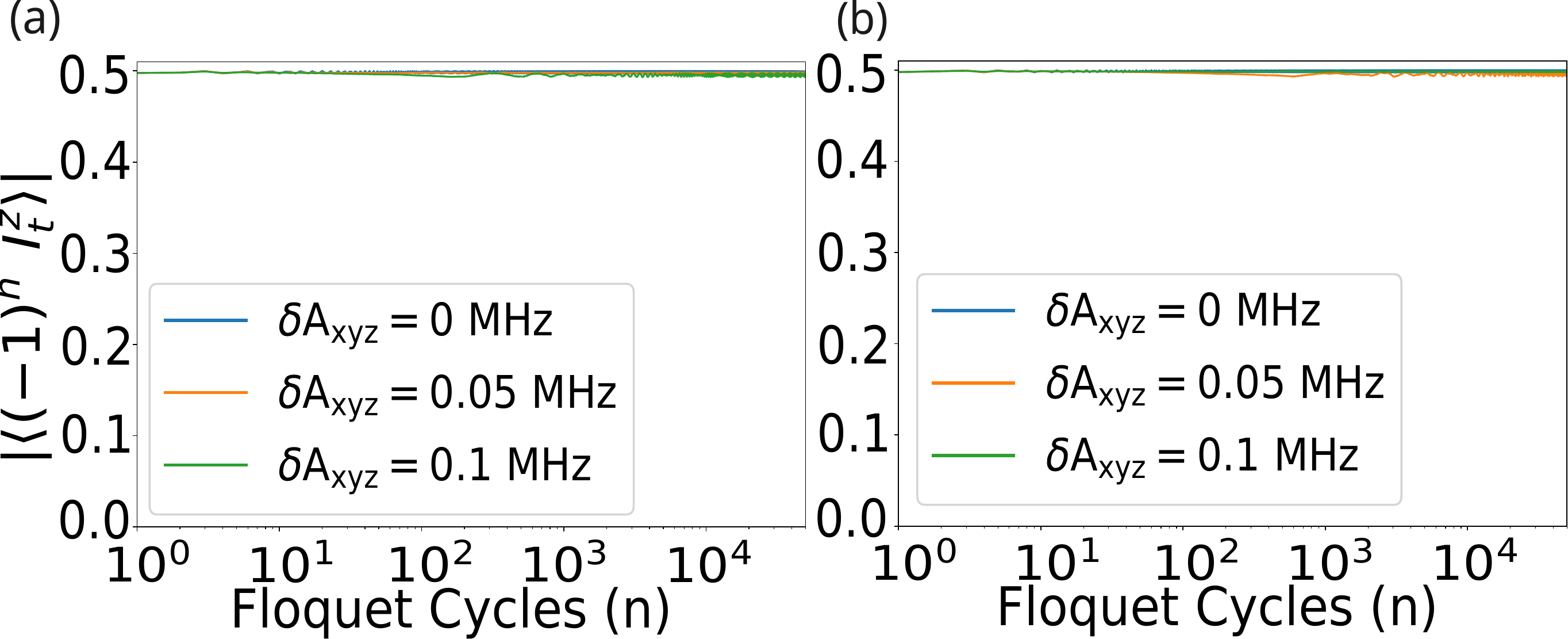}
  \caption{Effect of random couplings in the XXX and XXZ central-spin systems with a Zeeman field of $B_z=300$ MHz on the central spin.  The couplings are sampled from a Gaussian distribution with mean $A_{xyz}$ and variance (disorder strength) $\delta A_{xyz}$. (a) Staggered average satellite magnetization as a function of Floquet cycle for $A_{xyz}=1.3$ MHz for the XXX central-spin model. (b) Staggered average satellite magnetization as a function of Floquet cycle for $A_{xyz}=1.3$ MHz for the XXZ central-spin model, where $A_{z}=A_{xyz}\sqrt{0.9}$ and $A_{z}=A_{xyz}\sqrt{0.1}$. Here, $N=5$, the pulse error is $\theta=0.03\pi$, $\omega=1$ MHz, and the initial state is $\ket{J\downarrow}$.}
  \label{fig:fig10app}
\end{figure}

In Fig.~\ref{fig:fig10app} (a), (b), we show the staggered average magnetization for 
the total satellite spin $\I_t^z$ over many Floquet cycles for different coupling disorder strengths in the case of Heisenberg and XXZ  interactions, respectively. The average magnetization is close to 0.5 in both cases. This shows  that, for a fully polarized state, the time-crystalline behavior remains robust even in presence disorder coupling. In fact, this result is a special case of previous results in~\cite{Rafail_PRB_2023}, where an MBL-DTC has been shown to exist in disordered driven XXZ and XXZ models. 


\bibliography{References}
\end{document}